\documentclass[twocolumn]{aastex631}

\newcommand{\G}{\mathcal{G}}

\newcommand{\appropto}{\mathrel{\vcenter{\offinterlineskip\halign{\hfil$##$\cr\propto\cr\noalign{\kern2pt}\sim\cr\noalign{\kern-2pt}}}}}

\newcommand{\teff}{\ensuremath{T_{\rm eff}}}

\newcommand{\msun}{\ensuremath{\,{\rm M_\Sun}}}
\newcommand{\rsun}{\ensuremath{\,{\rm R_\Sun}}}

\newcommand{\mj}{\ensuremath{\,{\rm M_{\rm J}}}}
\newcommand{\rj}{\ensuremath{\,{\rm R_{\rm J}}}}
\newcommand{\degree}{\ensuremath{\,^{\circ}}}

\usepackage{url}
\usepackage{graphicx}
\usepackage{amsmath,amstext}
\usepackage[T1]{fontenc}

\usepackage{amsmath}
\usepackage{float}
\usepackage{longtable}
\usepackage[figuresright]{rotating}

\shorttitle{The PFS view of TOI-677 b}
\shortauthors{Hu et al.}

\begin{document}
\title{The PFS view of TOI-677 b: A spin-orbit aligned warm Jupiter in a dynamically hot system\footnote{This paper includes data gathered with the 6.5 meter Magellan Telescopes located at Las Campanas Observatory (LCO), Chile.}}

\author[0009-0007-9015-9451]{Qingru Hu} 
\affiliation{Weiyang College, Tsinghua University, Beijing, 100084, China}

\author[0000-0002-7670-670X]{Malena Rice} 
\affiliation{Department of Astronomy, Yale University, New Haven, CT 06511, USA}

\author[0000-0002-0376-6365]{Xian-Yu Wang} 
\affiliation{Department of Astronomy, Indiana University, Bloomington, IN 47405, USA}

\author[0000-0002-7846-6981]{Songhu Wang} 
\affiliation{Department of Astronomy, Indiana University, Bloomington, IN 47405, USA}

\author[0000-0002-1836-3120]{Avi~Shporer} 
\affiliation{Department of Physics and Kavli Institute for Astrophysics and Space Research, Massachusetts Institute of Technology, Cambridge, MA 02139, USA}

\author[0009-0008-2801-5040]{Johanna K. Teske} 
\affiliation{Carnegie Institution for Science, Earth \& Planets Laboratory, 5241 Broad Branch Road NW, Washington, DC 20015, USA}

\author[0000-0001-7961-3907]{Samuel W.\ Yee} 
\affiliation{Department of Astrophysical Sciences, Princeton University, 4 Ivy Lane, Princeton, NJ 08544, USA}
\affiliation{Center for Astrophysics \textbar \ Harvard \& Smithsonian, 60 Garden St, Cambridge, MA 02138, USA}
\altaffiliation{51 Pegasi b Fellow}

\author[0000-0003-1305-3761]{R. Paul Butler} 
\affiliation{Carnegie Institution for Science, Earth \& Planets Laboratory, 5241 Broad Branch Road NW, Washington, DC 20015, USA}

\author[0000-0002-8681-6136]{Stephen Shectman} 
\affiliation{The Observatories of the Carnegie Institution for Science, 813 Santa Barbara Street, Pasadena, CA 91101, USA}

\author[0000-0002-5226-787X]{Jeffrey D. Crane} 
\affiliation{The Observatories of the Carnegie Institution for Science, 813 Santa Barbara Street, Pasadena, CA 91101, USA}

\author[0000-0001-6588-9574]{Karen A.\ Collins} 
\affiliation{Center for Astrophysics \textbar \ Harvard \& Smithsonian, 60 Garden Street, Cambridge, MA 02138, USA}

\author[0000-0003-2781-3207]{Kevin I.\ Collins} 
\affiliation{George Mason University, 4400 University Drive, Fairfax, VA, 22030 USA}

\correspondingauthor{Qingru Hu}
\email{huqr20@mails.tsinghua.edu.cn}

\begin{abstract}

TOI-677 b is part of an emerging class of ``tidally-detached'' gas giants ($a/R_\star \gtrsim 11$) that exhibit large orbital eccentricities and yet low stellar obliquities. Such sources pose a challenge for models of giant planet formation, which must account for the excitation of high eccentricities without large changes in the orbital inclination. In this work, we present a new Rossiter-McLaughlin (RM) measurement for the tidally-detached warm Jupiter TOI-677 b, obtained using high-precision radial velocity observations from the PFS/Magellan spectrograph. Combined with previously published observations from the ESPRESSO/VLT spectrograph, we derive one of the most precisely constrained sky-projected spin-orbit angle measurements to date for an exoplanet. The combined fit offers a refined set of self-consistent parameters, including a low sky-projected stellar obliquity of $\lambda=3.2^{+1.6}_{-1.5}\degree$ and a moderately high eccentricity of $e=0.460^{+0.019}_{-0.018}$, that further constrains the puzzling architecture of this system. We examine several potential scenarios that may have produced the current TOI-677 orbital configuration, ultimately concluding that TOI-677 b most likely had its eccentricity excited through disk-planet interactions. This system adds to a growing population of aligned warm Jupiters on eccentric orbits around hot ($T_{\rm eff}>6100$ K) stars.

\end{abstract}

\vspace{-10mm}

\section{Introduction} 
\label{section:intro}

In contrast to the alignment of the solar spin and the planetary orbits in our solar system \citep{souami2012solar}, exoplanets have been discovered on inclined, polar \citep[e.g.][]{anderson2018wasp, addison2018polar, bourrier2020hot, albrecht2021preponderance, watanabe2022nodal} or even retrograde \citep[e.g.][]{neveu2014wasp, temple2017wasp, temple2019wasp} orbits around their host stars. The degree of the spin-orbit (mis)alignment of a planetary system is quantified by the "stellar obliquity", which is the angle between the planetary orbital angular momentum vector and the stellar spin angular momentum vector. The obliquity of a system is sculpted by both its degree of and mechanism for dynamical excitation, as well as subsequent tidal effects (see \citet{ogilvie2014tidal} for a review of tidal dissipation). Obliquity excitation mechanisms fall into three categories \citep{albrecht2022stellar}: primordial misalignment of the proto-planetary disk \citep{matsakos2017disk, takaishi2020diskalign, romanova2021MHD}, post-formation misalignment \citep{andreson2016kl, petrovich2016warm, anderson2018teetering, teyssandier2019formation, beauge2012planetscatter}, and changes in the stellar spin that are not related to planet formation \citep{rogers2012internal, rogers2013igw}. 

Stellar obliquities are commonly constrained using the Rossiter-McLaughlin effect \citep{Rossiter1924,McLaughlin1924}, which enables measurements of the sky-projected spin-orbit angle $\lambda$. A planet blocks out blue-shifted or red-shifted components of the rotating host star's light at different phases during its transit, resulting in a deviation from the overarching Doppler reflex motion in the host star's spectrum \citep{RM2000}. High-resolution spectroscopic observations across the transit can be leveraged to measure the Rossiter-McLaughlin effect.

To date, measurements of the stellar obliquity have been severely biased toward systems hosting a massive and close-orbiting ($a/R_\star \lesssim 11$) planet, which are referred to as ``hot Jupiters'', due to their frequent, short, and deep transits (see \cite{albrecht2022stellar} for a review of existing constraints). Only a handful of wider-orbiting ``warm Jupiter'' systems ($a/R_\star \gtrsim 11$) have precise spin-orbit constraints due to their low transit probabilities, long transit durations, and infrequent transit events. However, the orbital configurations of warm-Jupiter systems can play an important role in constraining the prevalence of competing planet formation and evolution pathways \citep{HJ_ARAA2015,dawson2018origins}.

A comparison between the orbital configurations of close-orbiting hot Jupiters and wide-orbiting warm Jupiters (such as the orbital eccentricity distribution, the companion rate, and so on) can shed light on the prevalence of different formation mechanisms for these two types of planets. The possible formation channels for hot and warm Jupiters are directly comparable: each class of planets could potentially arise through in situ formation, disk migration, or high-eccentricity migration. Previous work has shown that high-eccentricity tidal migration triggered by planet-planet or planet-star Kozai-Lidov cycles \citep[KL;][]{von_zeipel_1910, lidov1962evolution, kozai1962secular, wu2003planet, naoz2011hot, naoz2012formation, naoz2016eccentric} is a promising channel to account for the observed properties of hot Jupiters on a population level \citep{HJ_ARAA2015, dawson2018origins, HJ_obliquity_Rice_2022}. 

However, it is not yet clear how warm Jupiters fit into this emerging picture. Warm Jupiters, by contrast to hot Jupiters, show a high occurrence rate of nearby planetary companions \citep{huang2016warm, Wu_2023}. Furthermore, previous work has proposed that the warm Jupiter eccentricity distribution contains both a low-eccentricity component and a roughly uniform component \citep{dawson2018origins, petrovich2016warm}. In situ formation or disk migration could account for the low-eccentricity component of this distribution, producing warm Jupiters that cannot be excited by subsequent scattering \citep{petrovich2014cicularwj}. A separate mechanism is needed to reproduce the uniform eccentricity distribution component, which is also difficult to reproduce in the high-eccentricity migration framework as warm Jupiters are tidally detached from their host stars.

A comparison of the stellar obliquity distributions of hot and warm Jupiters, particularly around hot stars, can help to distinguish the prevalence of each mechanism by disentangling the stage at which misalignments arise \citep[][]{triaud2010hj, hamer2022evidence, attia2023DREAM}.
Hot Jupiters on tight orbits are strongly affected by tidal interactions with their host stars, such that the outer layer of the star may realign over time \citep{winn2010hot,albrecht2012obliquities}. Previous studies of hot Jupiter stellar obliquities exhibit evidence suggestive of tidal realignment \citep{albrecht2012obliquities}: hot stars hosting hot Jupiters span a wider range of obliquities than their cool star counterparts \citep{Schlaufman_2010, Winn_2010}. The transition occurs at the Kraft break ($\teff \sim 6100$ K), the dividing point for whether the stellar envelope is convection-dominated or radiation-dominated \citep{kraft1967studies}.

By contrast, warm Jupiters are expected to be ``tidally detached'' -- that is, they are relatively unaffected by tidal dissipation within the age of the system, such that they offer valuable insights into the primordial obliquity distribution at the time of planet formation \citep{zhou2020obliquity, rice2021soles}. 
\citet{WJ_obliquity_Rice_2022} suggests that small spin–orbit angles observed for warm Jupiters around single stars may indicate that protoplanetary disks tend to be aligned at the time of gas dispersal, while hot Jupiters are misaligned through dynamical interactions after the disk has dispersed. A larger sample of warm Jupiters with precisely measured spin-orbit orientations -- particularly those around hot stars, which have relatively few such measurements to date -- will enable us to examine the robustness of this preliminary statistical result and to further inform the key hot vs. warm Jupiter obliquity excitation mechanism(s).

The precise characterization of individual warm Jupiter orbital configurations is critical to inform competing models for the formation of warm Jupiters more generally. TOI-677 (TIC 280206394) was first identified as a planet-hosting candidate by the TESS mission \citep{ricker2015tess} and was then characterized by \citet{Jordan_2020} as a bright ($V=9.8$), hot ($\teff = 6295 \pm 77$ K) late-F star hosting a wide-orbiting eccentric ($e=0.435\pm0.024$) gas giant TOI-677 b. Recently, \citet{TOI677_Sedaghati_2023} reported a sky-projected obliquity measurement ($\lambda=0.3\pm1.3^{\circ}$) for this target over a single transit with the ESPRESSO \citep[Echelle Spectrograph for Rocky Exoplanets and Stable Spectroscopic Observations;][]{pepe2021espresso} spectrograph.

We present a new, independent Rossiter–McLaughlin measurement across one transit of TOI-677 b with the Carnegie Planet Finder Spectrograph \citep[PFS;][]{crane2006carnegie, crane2008carnegie, crane2010carnegie} on the 6.5m Magellan Clay telescope. This observation was obtained as part of the Stellar Obliquities in Long-period Exoplanet Systems (SOLES) survey \citep{rice2021soles, wang2022aligned, rice2022tendency, rice2023qatar6, hixenbaugh2023spin, dong2023toi, wright2023soles, rice2023evidence, lubin2023toi}, and it represents the tenth published result from this program.

We examine our new measurement, together with archival observations and data from \citet{Jordan_2020} and \citet{TOI677_Sedaghati_2023}, to refine the orbital parameters and the stellar obliquity of the TOI-677 system. By combining two independent RM datasets, each from high-precision spectrographs on 6-8m class telescopes, we derive one of the most precisely constrained sky-projected spin-orbit measurements to date: an important advance toward inferring precise orbital geometries and understanding the true dispersion of exoplanet orbital distributions. We find that TOI-677 b is well aligned with its host star, with $\lambda = {3.2^{+1.6}_{-1.5}}^{\circ}$. 

We describe the newly analyzed observations in Section \ref{sect:observations}. In Section \ref{sect:analysis}, we describe the joint analysis of photometry and RV data to get the projected stellar obliquity $\lambda$. We discuss the implications of this measurement in Section \ref{sect:discussion} and summarize our findings in Section \ref{sect:conclusion}.

\section{Observations}\label{sect:observations}

\subsection{PFS Observations}\label{subsectino:PFS}
We observed the Rossiter-McLaughlin effect across one full transit of TOI-677 b from UT 2:19-9:24 on February 10th, 2023 using the Carnegie Planet Finder Spectrograph \citep[PFS;][]{crane2006carnegie, crane2008carnegie, crane2010carnegie} on the 6.5m Magellan Clay telescope. In addition to the transit itself, our observation included 2.5 hours of pre-transit and 2 hours of post-transit baseline observations. Seeing was highly variable across the observing sequence, ranging from $1.0-2.5\arcsec$. The target was high in the sky, at airmass $z=1.18$ during the start of the observing sequence and reaching $z=1.71$ at the end. The moon was separated from the target by $60\degree$ throughout the night.

 The spectral data reduction and RV extraction were performed using a customized pipeline \citep{butler1996attaining}. The PFS radial velocity values and uncertainties used in this work are listed in Table \ref{tab:pfs_rv_data}.

\begin{deluxetable}{lccc}
\tablecaption{PFS radial velocities and uncertainties across a single transit of TOI-677\,b.\label{tab:pfs_rv_data}}
\tabletypesize{\scriptsize}
\tablehead{
\colhead{BJD (-2,450,000)} & \colhead{RV (m/s)} & \colhead{$\sigma_{\rm RV}$ (m/s)} }
\tablewidth{\linewidth}
\startdata
9985.604119 & 89.06 & 3.89 \\
9985.613910 & 90.47 & 3.48 \\
9985.623650 & 79.87 & 3.56 \\
9985.633890 & 65.70 & 3.29 \\
9985.643931 & 69.94 & 2.97 \\
9985.653941 & 63.16 & 2.82 \\
9985.664041 & 48.56 & 2.70 \\
9985.673402 & 46.57 & 2.90 \\
9985.683612 & 38.34 & 2.79 \\
9985.693662 & 30.57 & 2.51 \\
9985.703572 & 27.52 & 2.58 \\
9985.713363 & 23.82 & 2.36 \\
9985.723783 & 26.02 & 2.73 \\
9985.733683 & 16.15 & 2.66 \\
9985.743514 & 0.00 & 2.70 \\
9985.753874 & -9.34 & 2.52 \\
9985.763624 & -28.07 & 2.62 \\
9985.773135 & -46.97 & 3.17 \\
9985.783905 & -65.26 & 3.13 \\
9985.793885 & -75.57 & 3.31 \\
9985.803656 & -59.33 & 3.37 \\
9985.813456 & -66.81 & 3.46 \\
9985.823726 & -68.57 & 3.60 \\
9985.833226 & -66.60 & 3.91 \\
9985.843987 & -78.43 & 3.91 \\
9985.854247 & -92.25 & 3.36 \\
9985.863747 & -91.11 & 3.52 \\
9985.873578 & -99.67 & 3.49 \\
9985.883508 & -101.34 & 3.72 \\
\enddata
\tablecomments{This table is available in machine-readable form in the online manuscript.}
\end{deluxetable}

\subsection{TESS Photometry}\label{subsect:tess_photo}
TOI-677 (TIC 20182165) was observed at a two-minute cadence in TESS Sectors 9, 10, 35, 36, 62, and 63, including 12 transits in total. We downloaded the two-minute cadence SPOC TESS data from the Mikulski Archive for Space Telescopes (MAST) Portal \footnote{\url{https://mast.stsci.edu/portal/Mashup/Clients/Mast/Portal.html}}. We then detrended the TESS light curves to remove systematic trends by fitting a Gaussian Process (GP) model to the out-of-transit light curves using the \texttt{juliet} Python package \citep{Espinoza2019_Juliet}, which implements the Python package \texttt{celerite} \citep{Foreman-Mackey2017_celerite} to build the GP model. The detrended TESS data used for the analysis is presented in Table \ref{tab:tess_photo_data}. The phase-folded TESS light curve, along with the best-fit model from the combined fit in Table \ref{tab:results}, is shown in the leftmost panel of Figure \ref{fig:photo}.

\begin{deluxetable}{lccc}
\tablecaption{Detrended TESS photometric data for TOI-677. Relative flux values have been normalized to a baseline of 1.00.}\label{tab:tess_photo_data}
\tabletypesize{\scriptsize}
\tablehead{
\colhead{BJD (-2,450,000)} & \colhead{Flux (ppt)} & \colhead{$\sigma_{\rm flux}$ (ppt)}}
\tablewidth{\linewidth}
\startdata
8543.885982 & -0.938 & 0.819 \\
8543.887371 & -0.235 & 0.819 \\
8543.888760 & -2.335 & 0.818 \\
8543.890148 & -1.285 & 0.818 \\
8543.891537 & -1.345 & 0.818 \\
8543.892926 & -1.256 & 0.818 \\
8543.894315 & -1.257 & 0.818 \\
8543.895704 & -0.869 & 0.817 \\
8543.897093 & -1.712 & 0.817 \\
8543.898482 & -0.338 & 0.817 \\
8543.899871 & 0.401 & 0.817 \\
8543.901260 & -0.503 & 0.817 \\
8543.902649 & -1.412 & 0.816 \\
8543.904038 & -0.585 & 0.816 \\
8543.905426 & 0.500 & 0.816 \\
... & ... & ... \\
\enddata
\tablecomments{This table is available in machine-readable form in the online manuscript.}
\end{deluxetable}

\subsection{LCO Ground-based Photometry}\label{subsectino:GB_Photo}
\begin{figure*}
    \centering
    \includegraphics[width=\linewidth]{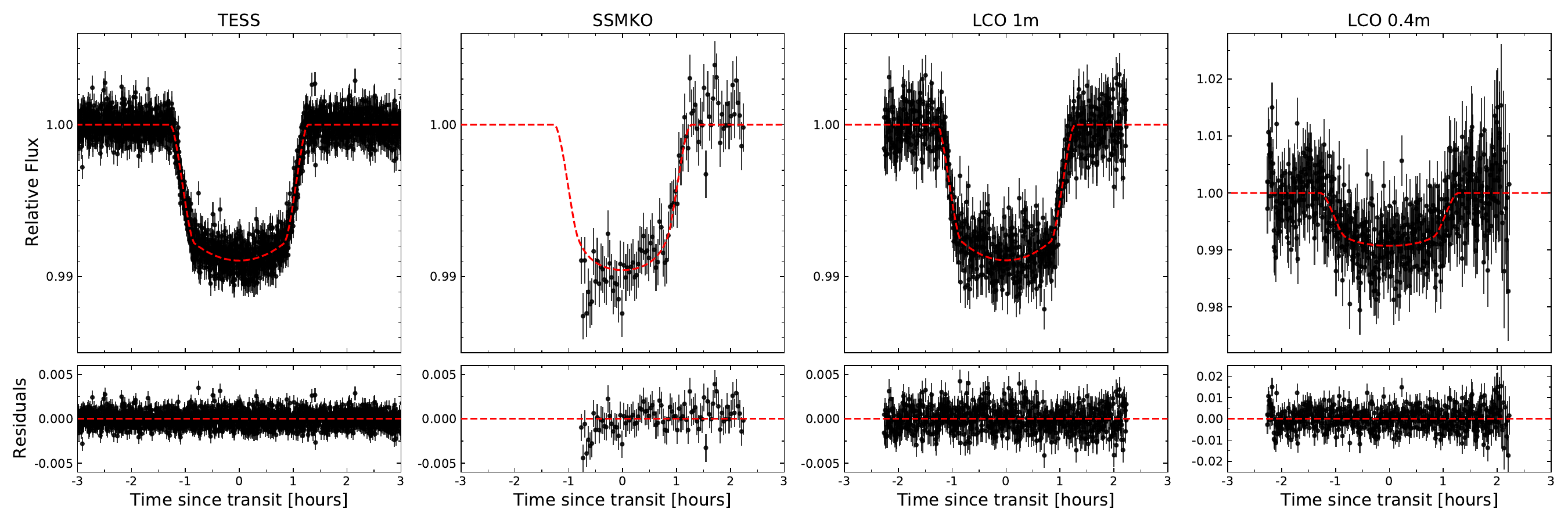}
    \caption{Phase-folded light curves for TESS, SSMKO, LCO 1m and LCO 0.4m, with the overplotted best-fit models from our combined fit as red dashed lines. The corresponding parameters are listed in Table \ref{tab:results}.}
    \label{fig:photo}
\end{figure*}

We obtained photometry of TOI-677 that was simultaneous with the RM observation using two 1m and two 0.4m telescopes within the Las Cumbres Observatory Global Telescope \citep[LCOGT;][]{Brown:2013} network on UT February 10th, 2023. All LCOGT observations were obtained with the Pan-STARRS $zs$ filter.\footnote{\url{https://lco.global/observatory/instruments/filters/PanSTARRS-zs/}} One of the 1m telescopes did not successfully defocus and nearly saturated the target star; however, a reasonable lightcurve could still be extracted. The second 1m telescope, as well as the two 0.4m telescopes, successfully defocused with no problems. The images were calibrated by the standard LCOGT {\tt BANZAI} pipeline \citep{McCully:2018} and differential photometric data were extracted using {\tt AstroImageJ} \citep{Collins:2017}. The detrended LCO data used for the analysis are presented in Table \ref{tab:lco_photo_data}. We combined photometry from telescopes of the same diameter into one dataset (labeled as ``LCO 1m'' and ``LCO 0.4m'') in our analysis. The phase-folded LCO light curves from the 1m and 0.4m telescopes, along with the best-fitting model from the combined fit (see Table \ref{tab:results}), are shown in the third and fourth panels in Figure \ref{fig:photo}.

\begin{deluxetable}{lccc}
\tablecaption{LCO photometric data for the TOI-677 b transits examined in this work. Relative flux values have been normalized to a baseline of 1.00. The two 0.4m telescopes and the two 1m telescopes are distinguished with the markers (1) and (2). \label{tab:lco_photo_data}}
\tabletypesize{\scriptsize}
\tablehead{
\colhead{BJD (-2,450,000)} & \colhead{Flux (ppt)} & \colhead{$\sigma_{\rm flux}$ (ppt)} & \colhead{Telescope}}
\tablewidth{\linewidth}
\startdata
9985.664765 & 1.544 & 3.481 & LCO 0.4m (1) \\
9985.665403 & 1.483 & 3.483 & LCO 0.4m (1) \\
9985.666047 & 8.032 & 3.488 & LCO 0.4m (1) \\
9985.666682 & -0.989 & 3.453 & LCO 0.4m (1) \\
9985.667324 & 6.566 & 3.482 & LCO 0.4m (1) \\
...&...&...&...\\
9985.664785 & -3.179 & 3.383 & LCO 0.4m (2) \\
9985.665528 & 10.224 & 3.378 & LCO 0.4m (2) \\
9985.666258 & 6.324 & 3.392 & LCO 0.4m (2) \\
9985.666981 & 5.986 & 3.390 & LCO 0.4m (2) \\
9985.667688 & 4.987 & 3.378 & LCO 0.4m (2) \\
...&...&...&...\\
9985.664984 & 0.617 & 0.713 & LCO 1m (1) \\
9985.665619 & 1.120 & 0.711 & LCO 1m (1) \\
9985.666269 & 0.606 & 0.710 & LCO 1m (1) \\
9985.666856 & 0.221 & 0.710 & LCO 1m (1) \\
9985.667443 & 0.213 & 0.711 & LCO 1m (1) \\
...&...&...&...\\
9985.664894 & -1.433 & 0.698 & LCO 1m (2) \\
9985.665489 & -1.568 & 0.698 & LCO 1m (2) \\
9985.666078 & -1.877 & 0.698 & LCO 1m (2) \\
9985.666671 & -1.811 & 0.698 & LCO 1m (2) \\
9985.667260 & -1.793 & 0.699 & LCO 1m (2) \\
\enddata
\tablecomments{The table is in machine-readable form in the online manuscript.}
\end{deluxetable}

\section{Analysis and Results}\label{sect:analysis}

\begin{deluxetable*}{lcccc}
\tablewidth{\linewidth}
\tabletypesize{\scriptsize}
\tablecaption{Parameters, Priors, and Results for TOI-677 b.  \label{tab:results}}
\tablehead{
\colhead{Parameter} & \colhead{Priors} & \colhead{PFS} & \colhead{ESPRESSO} & \colhead{Combined Fit}}
\startdata
\textit{Fitted Parameters:}\\
$R_{\rm b} / R_\star$&  $\mathcal U(0.0942;0;1)$   & $0.09429^{+0.00050}_{-0.00065}$ & $0.09404^{+0.00049}_{-0.00064}$ & 
$0.09404^{+0.00049}_{-0.00064}$ \\
$(R_\star + R_{\rm b}) / a_{\rm b}$ & $\mathcal U(0.0627;0;1)$ &  $0.0623^{+0.0017}_{-0.0014}$  & $0.0623^{+0.0015}_{-0.0013}$ & 
$0.0625 \pm -0.0014$ \\
$\cos{i_{\rm b}}$ &    $\mathcal U(0.0413;0;1)$   & $0.0726^{+0.0049}_{-0.0041}$    & $0.0729^{+0.0044}_{-0.0037}$ & 
$0.0738^{+0.0042}_{-0.0039}$ \\
$T_{\rm 0, b}$ (BJD-2458500)& $\mathcal U(47.4744;46.4744;48.4744)$ & $47.47439 \pm 0.00016$ & $47.47448\pm 0.00016$ & 
$47.47449 \pm 0.00016$ \\
$P_{\rm b}$ (day)& $\mathcal U(11.2366;10;12)$   & $11.2365991 \pm 0.0000017$ & $11.2365986 \pm 0.0000019$ & 
$11.2365985 \pm 0.0000019$ \\
$K_{\rm b}$ (m/s)&  $\mathcal U(113.9;0;10000)$   & $112.3^{+4.1}_{-3.9}$ & $112.3^{+3.9}_{-4.3}$ & 
$112.9^{+4.0}_{-4.1}$ \\
$\sqrt{e_{\rm b}} \cos{\omega_{\rm b}}$&   $\mathcal U(0.2204;-1;1)$   &  $0.196^{+0.039}_{-0.036}$ & $0.199^{+0.040}_{-0.036}$ & 
$0.199^{+0.039}_{-0.036}$ \\
$\sqrt{e_{\rm b}} \sin{\omega_{\rm b}}$  &   $\mathcal U(0.6216;-1;1)$   &  $0.641^{+0.019}_{-0.018}$  & $0.644^{+0.017}_{-0.016}$ & 
$0.647 \pm 0.016$ \\
$\lambda \ (^{\circ})$    &  $\mathcal U(0;-180;180)$   & $7.9^{+2.4}_{-2.2}$   & $0.9^{+1.6}_{-1.5}$ & 
$3.2^{+1.6}_{-1.5}$ \\
$v \sin i_\star$ (km/s)& $\mathcal U(7.80;0;20)$   & $6.58^{+0.78}_{-0.64}$  & $8.57^{+0.83}_{-0.69}$ & 
$7.88^{+0.60}_{-0.53}$ \\
$q_{1;\text{TESS}}$ & $\mathcal U(0.5;0;1)$ & $0.14^{+0.04}_{-0.02}$ & $0.14_{-0.02}^{+0.03}$ & $0.14_{-0.02}^{+0.03}$\\
$q_{2;\text{TESS}}$ & $\mathcal U(0.5;0;1)$ & $0.72^{+0.25}_{-0.19}$ & $0.76_{-0.24}^{+0.16}$ & $0.77_{-0.24}^{+0.16}$\\
$q_{1;\text{SSMKO}}$ & $\mathcal U(0.5;0;1)$ & $0.61^{+0.23}_{-0.24}$ & $0.63_{-0.24}^{+0.22}$ & $0.62_{-0.24}^{+0.23}$\\
$q_{2;\text{SSMKO}}$ & $\mathcal U(0.5;0;1)$ & $0.06^{+0.10}_{-0.05}$ & $0.06_{-0.05}^{+0.10}$ & $0.06_{-0.05}^{+0.11}$\\
$q_{1;\text{LCO 1m}}$ & $\mathcal U(0.5;0;1)$ & $0.17^{+0.05}_{-0.04}$ & $0.16_{-0.04}^{+0.05}$ & $0.16_{-0.04}^{+0.05}$\\
$q_{2;\text{LCO 1m}}$ & $\mathcal U(0.5;0;1)$ & $0.79^{+0.14}_{-0.23}$ & $0.78_{-0.24}^{+0.16}$ & $0.78_{-0.25}^{+0.16}$\\
$q_{1;\text{LCO 0.4m}}$ & $\mathcal U(0.5;0;1)$ & $0.38^{+0.23}_{-0.18}$ & $0.37_{-0.17}^{+0.23}$ & $0.37_{-0.17}^{+0.24}$\\
$q_{2;\text{LCO 0.4m}}$ & $\mathcal U(0.5;0;1)$ & $0.38^{+0.35}_{-0.25}$ & $0.39_{-0.26}^{+0.34}$ & $0.39_{-0.26}^{+0.34}$\\
$q_{1;\text{PFS}}$ & $\mathcal U(0.5;0;1)$ & $0.62^{+0.23}_{-0.26}$ & ... & $0.75_{-0.17}^{+0.16}$\\
$q_{2;\text{PFS}}$ & $\mathcal U(0.5;0;1)$ & $0.39^{+0.36}_{-0.27}$ & ... & $0.71_{-0.26}^{+0.20}$\\
$q_{1;\text{ESPRESSO}}$ & $\mathcal U(0.5;0;1)$ & ... & $0.70^{+0.17}_{-0.16}$ & $0.67_{-0.20}^{+0.19}$\\
$q_{2;\text{ESPRESSO}}$  & $\mathcal U(0.5;0;1)$ & ... & $0.50_{-0.31}^{+0.32}$ & $0.28_{-0.20}^{+0.33}$\\ 
\\
\textit{Stellar Parameter Priors:}\\
$R_{\star} \ (\rsun)$ & $\mathcal{G}(1.28;0.03)$ & ... & ... & ... \\
$M_{\star} \ (\rsun)$ & $\mathcal{G}(1.181;0.058)$ & ... & ... & ... \\
\\
\textit{Derived Parameters:}\\
$R_{\rm b} \ (\rj)$ &    ... & $1.174 \pm 0.029 $ & $1.170 \pm 0.029$ & 
$1.170^{+0.029}_{-0.030}$ \\
$M_{\rm b} \ (\mj)$  &   ... & $1.232^{+0.088}_{-0.082}$ & $1.225^{+0.083}_{-0.082}$ & 
$1.234^{+0.084}_{-0.083}$ \\
$a_{\rm b}/R_\star$ & ... & $17.55^{+0.42}_{-0.47}$ & $17.57^{+0.38}_{-0.42}$ & 
$17.49^{+0.39}_{-0.40}$ \\
$b$ & ... & $0.7125^{+0.0098}_{-0.0112}$ & $0.709^{+0.010}_{-0.012}$ & 
$0.709^{+0.010}_{-0.012}$ \\
$T_{\rm 14}$ (hours) & ... & $2.544 \pm 0.011$ & $2.536^{+0.012}_{-0.011}$ & 
$2.535 \pm 0.011$ \\
$i_{\rm b} \ (^{\circ})$ &  ... & $85.83^{+0.23}_{-0.28}$ & $85.82^{+0.21}_{-0.25}$ & 
$85.77^{+0.22}_{-0.24}$ \\          
$e_{\rm b}$ &   ... & $0.451^{+0.023}_{-0.021}$ & $0.455^{+0.020}_{-0.018}$ & 
$0.460^{+0.019}_{-0.018}$ \\
$\omega_{\rm b} \ (^{\circ})$ & ...& $73.1^{+3.1}_{-3.4}$ & $72.8^{+3.2}_{-3.4}$ & 
$72.9^{+3.2}_{-3.4}$ \\
$a_{\rm b}$ (au) & ...& $0.1044^{+0.0036}_{-0.0039}$ & $0.1045^{+0.0034}_{-0.0036}$ & 
$0.1041^{+0.0035}_{-0.0034}$  \\
$u_{1;\text{TESS}}$ & ... & $0.53_{-0.15}^{+0.10}$ & $0.56_{-0.14}^{+0.09}$ & $0.57_{-0.15}^{+0.09}$\\
$u_{2;\text{TESS}}$ & ... & $-0.16_{-0.12}^{+0.19}$ & $-0.19_{-0.10}^{+0.18}$ & $-0.20_{-0.10}^{+0.18}$\\
$u_{1;\text{SSMKO}}$ & ... & $0.09_{-0.07}^{+0.14}$ & $0.10_{-0.07}^{+0.15}$ & $0.10_{-0.07}^{+0.15}$\\
$u_{2;\text{SSMKO}}$ & ... & $0.67_{-0.22}^{+0.17}$ & $0.67_{-0.22}^{+0.17}$ & $0.66_{-0.23}^{+0.17}$\\
$u_{1;\text{LCO 1m}}$ & ... & $0.63_{-0.16}^{+0.11}$ & $0.61_{-0.16}^{+0.12}$ & $0.60_{-0.17}^{+0.12}$\\
$u_{2;\text{LCO 1m}}$ & ... & $-0.24_{-0.10}^{+0.18}$ & $-0.22_{-0.11}^{+0.19}$ & $-0.22_{-0.11}^{+0.19}$\\
$u_{1;\text{LCO 0.4m}}$ & ... & $0.45_{-0.29}^{+0.35}$ & $0.45_{-0.29}^{+0.34}$ & $0.45_{-0.30}^{+0.35}$\\
$u_{2;\text{LCO 0.4m}}$ & ... & $0.14_{-0.37}^{+0.36}$ & $0.12_{-0.35}^{+0.36}$ & $0.13_\pm0.36$\\
$u_{1;\text{PFS}}$   & ... & $0.58_{-0.40}^{+0.51}$ & ... & $1.20_{-0.42}^{+0.29}$\\
$u_{2;\text{PFS}}$   & ... & $0.16_{-0.50}^{+0.44}$ & ... & $-0.36_{-0.30}^{+0.46}$\\
$u_{1;\text{ESPRESSO}}$   & ... & ... & $0.83_{-0.51}^{+0.45}$ & $0.45_{-0.32}^{+0.47}$\\
$u_{2;\text{ESPRESSO}}$   & ... & ... & $0.00_{-0.50}^{+0.54}$ & $0.36_{-0.51}^{+0.35}$\\
\enddata
\end{deluxetable*}

\begin{figure*}
    \centering
    \includegraphics[width=\linewidth]{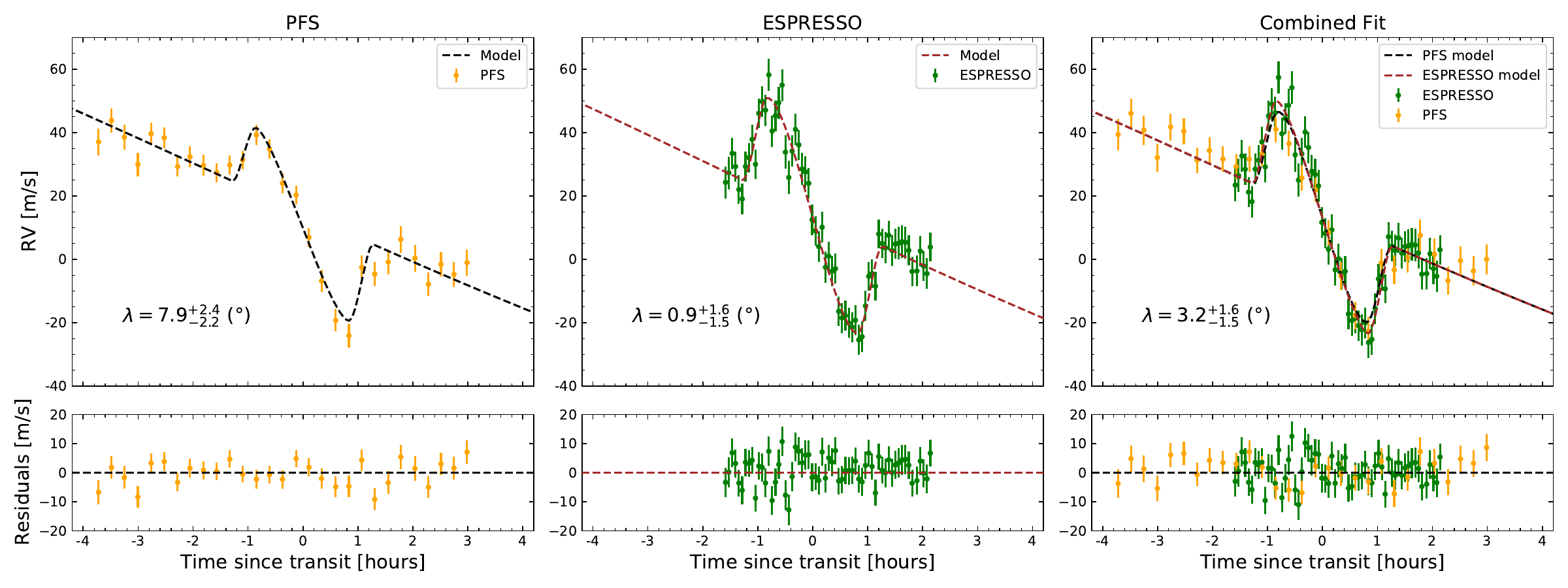}
    \caption{In-transit radial velocities and the best-fit models for our PFS, ESPRESSO, and combined datasets, with results corresponding to the three models described in Table \ref{tab:results}. The orange and green points label the PFS and ESPRESSO data, respectively. The PFS model is shown as a black dashed line, and the ESPRESSO model is shown as a brown dashed line. Quadratic limb-darkening coefficients were modeled separately for PFS and ESPRESSO as listed in Table \ref{tab:results}, so there are two overplotted model curves in the right panel. The RV baseline is subtracted from the radial velocities, and jitter terms have been added in quadrature to the RV uncertainties.}
    \label{fig:RM}
\end{figure*}

We carried out three independent joint fits using in-transit RV data only from PFS, only from ESPRESSO, and from both telescopes together. Each of these fits included photometric data from TESS and LCO (see Section \ref{sect:observations}), as well as archival photometric data from SSMKO (see the second panel in Figure \ref{fig:photo}) and archival RV data from the FEROS, Coralie, CHIRON, NRES, and \textsc{Minerva}-Australis spectrographs. The in-transit PFS RV data are provided in Table \ref{tab:pfs_rv_data}, and the in-transit ESPRESSO RV data are adopted from Table 4 in \citet{TOI677_Sedaghati_2023}. The SSMKO photometry data and all other archival RV data are drawn from Tables 2 and 4 in \citet{Jordan_2020}.

Our fits each used \texttt{allesfitter} \citep{gunther2021allesfitter}, which is a powerful tool for jointly modeling photometric and RV data. \texttt{Allesfitter} constructs a Bayesian inference framework that unites the \texttt{ellc} (light curve and RV models; \cite{maxted2016ellc}) and  \texttt{emcee} (Markov Chain Monte Carlo (MCMC) sampling; \cite{emcee}) Python packages, building a joint model of photometry and radial velocities to fit the Rossiter-McLaughlin effect.

All fitted parameters listed in Table \ref{tab:results} were allowed to vary and were initialized with uniform priors $\mathcal U(a; b)$, with a lower bound $a$ and an upper bound $b$. $R_{\rm b} / R_\star$ is the planet-to-star radius ratio; $(R_\star + R_{\rm b}) / a_{\rm b}$ is the sum of radii divided by orbital semimajor axis; $\cos{i_{\rm b}}$ is the cosine of the orbital inclination; $T_{\rm 0;b}$ is the mid-transit epoch; $P_{\rm b}$ is the orbital period; $K_{\rm b}$ is the RV semi-amplitude; $\sqrt{e}\cos{\omega}_{\rm b}$ and $\sqrt{e}\sin{\omega}_{\rm b}$ are the eccentricity parameters; $v \sin i_\star$ is the stellar rotation velocity projected on the line of sight; $\lambda$ is the sky-projected stellar obliquity; $q_1$ and $q_2$ are the quadratic limb darkening coefficients \citep{kipping2013ld} treated separately for all instruments that captured transit signals (TESS, SSMKO, LCO 1m, LCO 0.4m, PFS, and ESPRESSO). The initial values for $R_{\rm b} / R_\star$, $(R_\star + R_{\rm b}) / a_{\rm b}$, $\cos{i_{\rm b}}$, $T_{\rm 0;b}$, $P_{\rm b}$, $K_{\rm b}$, $\sqrt{e}\cos{\omega}_{\rm b}$, and $\sqrt{e}\sin{\omega}_{\rm b}$ were adopted from \citet{Jordan_2020}.

The sky-projected spin-orbit angle $\lambda$ was initialized as $0^{\circ}$ and was allowed to vary between $-180^{\circ}$ and $180^{\circ}$. The two limb-darkening coefficients $q_1$ and $q_2$ were initialized with values of 0.5 and were only allowed to vary between 0 and 1. Jitter terms were modeled separately for each instrument and were added in quadrature to the formal uncertainties. The baseline model was evaluated separately for photometric and radial velocity data. We allowed for a constant offset with priors bounded between -0.1 and 0.1 in relative flux units for photometric data. For RV data, we adopted a linear baseline model, as previous work has identified a significant long-term trend in the radial velocities which could be caused by an outer companion \citep{Jordan_2020}. We assigned a general linear slope for the archival radial velocities and different linear slopes for the PFS and ESPRESSO datasets. The constant RV offsets were different for individual spectrographs and were allowed to vary between -1 km/s and 1 km/s.

Stellar parameter priors are needed for the derived parameters, both listed in Table \ref{tab:results}. We adopted Gaussian priors for the host star's mass and radius drawn from \cite{Jordan_2020}, where they were derived from a combination of spectroscopy, spectral energy distribution (SED), and isochrone fits. The derived parameters are the planetary radius ($R_{\rm b}$), planetary mass ($M_{\rm b}$), planetary semi-major axis over host star radius ($a_{\rm b}/R_\star$), impact parameter ($b$), total transit duration ($T_{\rm 14}$), inclination of the planetary orbit ($i_{\rm b}$), eccentricity ($e_{\rm b}$), argument of periastron ($\omega_{\rm b}$), semi-major axis ($a_{\rm b}$), and the re-parameterized quadratic limb darkening coefficients $u_1$ and $u_2$ for TESS, SSMKO, LCO 1m, LCO 0.4m, PFS, and ESPRESSO.

We ran an affine-invariant MCMC analysis for each of our three joint fits, with 150 walkers to sample the posterior distributions of all model parameters. We ensured that each chain was fully converged by running all Markov chains to over 30 times their autocorrelation length, which corresponded to a minimum of 250,000 accepted steps per walker.

The values and 1-$\sigma$ uncertainties of fitted parameters are listed in Table \ref{tab:results}. The best-fit joint models are shown in Figure \ref{fig:RM}, together with the values and uncertainties for each $\lambda$ measurement. The jitter term for each instrument has been incorporated into the error bar of the plotted data points.

The sky-projected spin-orbit angles derived from PFS, ESPRESSO, and the combined fit are $\lambda=7.9^{+2.4}_{-2.2}\degree$, $\lambda=0.9^{+1.6}_{-1.5}\degree$, and $\lambda=3.2^{+1.6}_{-1.5}\degree$ respectively. These values are consistent with each other within 2-$\sigma$, and our final, combined fit measurement is also consistent with the result from \citet{TOI677_Sedaghati_2023} within 2-$\sigma$. TESS out-of-transit light curves of TOI-677 are nearly flat and bear no significant periodicities that could be attributed to the stellar spin, preventing a further derivation of the true 3D spin-orbit angle $\psi$. 

While the PFS measurement alone is suggestive of a solar-system-like stellar obliquity ($6\degree$ away from alignment; \citet{souami2012solar}), the ESPRESSO measurement alone is more indicative of exact alignment. Our final result, which combines the two, lies at an intermediate value between these two possibilities. The 2-$\sigma$ discrepancy between $\lambda$ values from the PFS and ESPRESSO measurements is discussed in detail in Section \ref{subsection:diff}.

\section{Discussion}\label{sect:discussion}

\subsection{Potential formation channels}\label{subsect:wjformation}
TOI-677 b displays a surprising orbital geometry: while it is a warm Jupiter on a substantially eccentric orbit ($e\sim0.46$) around a hot star \citep[$T_{\rm eff}=6295\pm77$ K;][]{Jordan_2020}, suggesting potential dynamical upheaval, the planet's orbit is precisely aligned with the host star's spin in its sky-plane projection ($\lambda < 10^{\circ}$). 
One possibility is that TOI-677 b may have had its orbital eccentricity excited through planet-disk interactions. Alternatively, the long-term linear RV trend followed by TOI-677 indicates the likely presence of an unconfirmed outer brown dwarf companion in the system (see Table 3 in \citet{TOI677_Sedaghati_2023}), which may have played a role in exciting TOI-677 b's eccentricity. The Coplanar High-Eccentricity Migration \citep[CHEM,][]{petrovich2015CHEM} mechanism or Kozai-Lidov (KL) oscillations \citep{fabrycky2007shrinking, naoz2011hot, naoz2012formation} may each potentially account for the formation of aligned ($\lambda<10^{\circ}$), moderately eccentric ($e\lesssim0.9$) warm Jupiters with an outer companion.

In the following subsections, we discuss each of these possible formation channels and the conditions under which they are effective. Ultimately, we find that the CHEM and KL mechanisms are unlikely to produce the observed system configuration, leaving planet-disk interactions as the leading hypothesis. The properties of the companion brown dwarf may offer further evidence to distinguish between potential formation channels for TOI-677 b.

\subsubsection{Coplanar High-Eccentricity Migration}
Coplanar high-eccentricity migration, or CHEM \citep{petrovich2015CHEM, zink2023hot}, offers one avenue to produce highly eccentric warm Jupiters within the plane of the natal protoplanetary disk. In the CHEM framework, a moderately or highly eccentric outer companion can excite the inner planet's eccentricity in-situ when the mutual inclination between the planet's orbit and the companion's orbit is small ($i_{\rm mut} \lesssim 20^{\circ}$). 

Following the arguments of \citet{TOI677_Sedaghati_2023}, the measured eccentricity $e_{\rm c} =0.436^{+0.067}_{-0.058}$ of the companion fails to excite the eccentricity of the inner planet to the new value $e_{\rm p}=0.460^{+0.019}_{-0.018}$ from our combined fit. Therefore, the CHEM mechanism would struggle to account for TOI-677 b's configuration unless the companion's eccentricity is found to be substantially higher than this previous estimate.

\subsubsection{Formation through the KL mechanism followed by tidal dissipation}
The Kozai-Lidov (KL) mechanism followed by tidal dissipation offers another potential avenue to produce aligned and substantially eccentric warm Jupiters. Hierarchical three-body systems, where the ratio of inner and outer orbital semimajor axes is much smaller than unity, may undergo KL oscillations over secular timescales. During these oscillations, the mutual inclination between the inner and outer orbits is traded back and forth with the inner orbit's eccentricity such that the component of the inner planet's orbital angular momentum parallel to that of the perturber ($L_{\rm z}$) is conserved as \citep[e.g.][]{tremaine2023dynamics}
\begin{equation}\label{equ:Lz}
\begin{aligned}
    L_{\rm z} = \sqrt{GM_\star a_{\rm p}(1-e_{\rm p}^2)} \cos i_{\rm mut} &= \text{constant} \\
    \sqrt{(1-e_{\rm p}^2)}\cos i_{\rm mut} &= \text{constant},
\end{aligned}
\end{equation}
where $M_\star$ is the mass of the host star, $a_{\rm p}$ is the semimajor axis of the planet (which remains unchanged due to energy conservation), $e_{\rm p}$ is the eccentricity of the planet, and $i_{\rm mut}$ is the mutual inclination between the planet's and the perturber's orbital planes. 

From Equation \ref{equ:Lz}, we can see that the KL mechanism offers a pathway to produce a considerable increase in the inner planet's eccentricity in exchange for a small change in the system's mutual inclination $i_{\rm mut}$. Under the assumption that planets generally form in protoplanetary disks well aligned with their host stars, small changes in the mutual inclination $\Delta i_{\rm mut}$ are equivalent to small stellar obliquities $\psi$. If the initial mutual inclination $i_{\rm mut,0}$ is larger, a smaller change in the mutual inclination is needed to excite the eccentricity from 0 to a set value through the KL mechanism. 

\begin{table}
\centering
    \caption{The current observed orbital configuration of TOI-677 adopted for calculation in this section. The parameters of the inner planet, TOI-677 b, are taken from the joint-fit results in Section \ref{sect:analysis}. The parameters of the outer companion are taken from the eccentric-orbit results from Table 3 in \citet{TOI677_Sedaghati_2023}. The parameters of the host star are taken from Table 1 in \citet{TOI677_Sedaghati_2023}.}
    \label{tab:orbit}
    \begin{tabular}{lcc}
    \hline\hline
    Description & Parameter & Value \\
    \hline
    \textit{The inner planet} &&\\
    \hline
    Eccentricity & $e_{\rm p}$ & 0.460 \\
    Orbital period & $P_{\rm p}$ (day) & 11.2365985 \\
    Semi-major axis & $a_{\rm p}$ (au) & 0.1041 \\
    Argument of periapsis & $\omega_{\rm p} \ (^{\circ})$ & 72.9 \\
    Mass & $M_{\rm p} \ (\mj)$ & 1.234 \\
    Radius & $R_{\rm p} \ (\rj)$ & 1.170\\
    \hline
    \textit{The outer companion} &&\\
    \hline
    Eccentricity &  $e_{\rm c}$ & 0.436 \\
    Orbital period & $P_{\rm c}$ (day) & 4901.13876 \\
    Semi-major axis & $a_{\rm c}$ (au) & 5.9872 \\
    Argument of periapsis & $\omega_{\rm c} \ (^{\circ})$ & -123.68 \\
    Mass & $M_{\rm c} \ (\mj)$ & 49.99 \\
    \hline
    \textit{The host star} &&\\
    \hline
    Mass & $M_\star \ (\msun)$ & 1.158 \\
    Radius & $R_\star \ (\rsun)$ & 1.281 \\
    \hline\hline
    \end{tabular}
\end{table}

\begin{figure*}
    \centering
    \includegraphics[width=\textwidth]{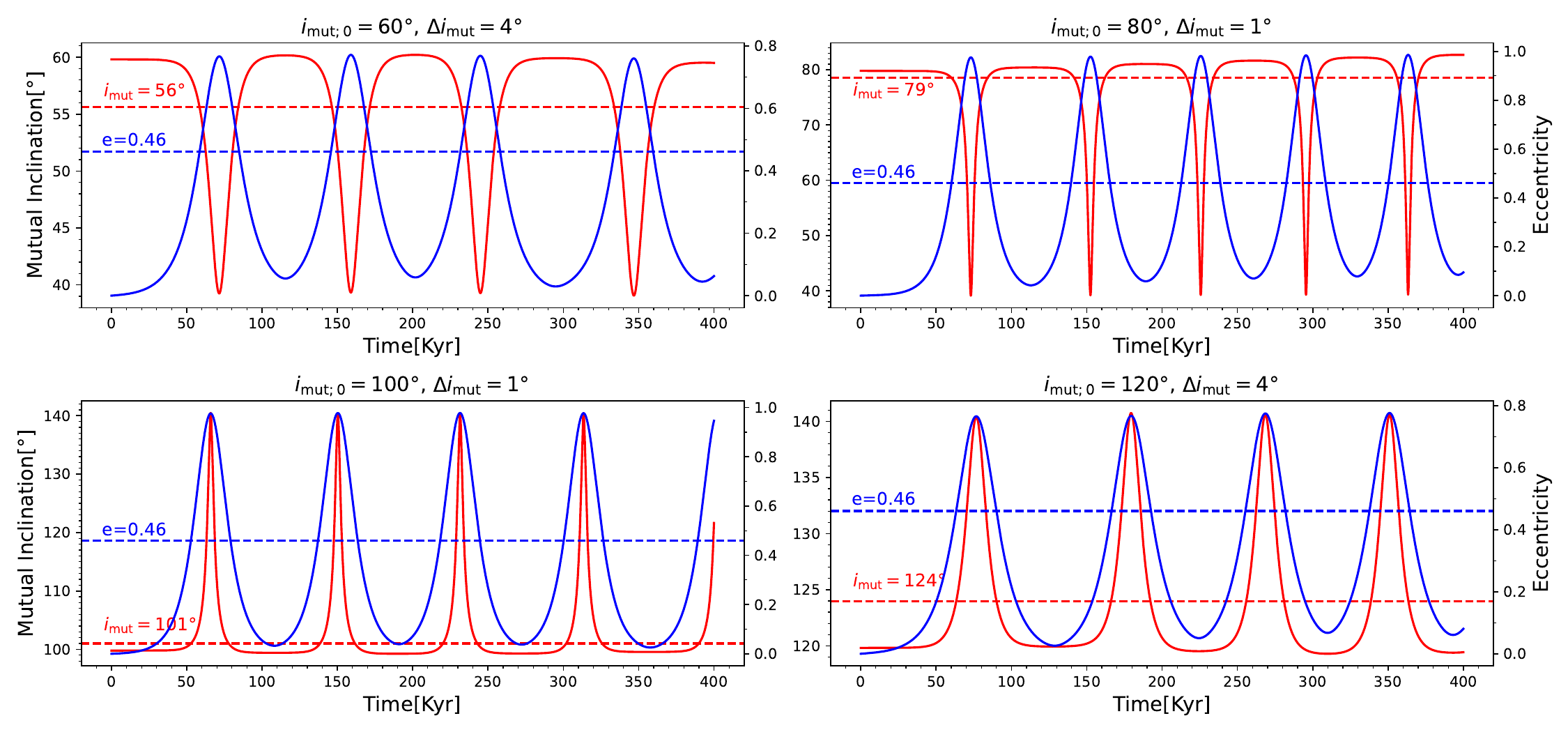}
    \caption{The \texttt{Kozaipy} simulation results for different starting mutual inclinations $i_{\rm mut;0}$ using initial conditions listed in Table \ref{tab:orbit} except that the eccentricity of the inner planet is set to nearly zero ($e_{\rm p}\approx0$). The mutual inclination oscillation is shown in red, while the eccentricity oscillation is in blue. The blue dashed lines show when the eccentricity oscillation reaches the currently observed value for TOI-677 b, and the red dashed lines demarcate the mutual inclinations at those times.}
    \label{fig:kozai}
\end{figure*}

We conducted numerical integrations for the inner planet's orbit using \texttt{Kozaipy}\footnote{\url{https://github.com/djmunoz/kozaipy}} to confirm our argument. The starting orbital configurations for the simulations, except that the eccentricity of the inner planet is set to nearly zero ($e_{\rm p}\approx0$), are the same as the current observed orbital configurations listed in Table \ref{tab:orbit} . We ran the simulation for four different initial mutual inclinations $i_{\rm mut;0}=60^{\circ}, 80^{\circ}, 100^{\circ}$, and $120^{\circ}$. Our results are shown in Figure \ref{fig:kozai}, with the eccentricity oscillation shown in blue and the mutual inclination oscillation shown in red. The blue dashed line marks the current, observed eccentricity (0.46) of the inner planet, while the red dashed line marks the mutual inclinations when the eccentricity reaches the observed value in the KL cycles. For initial values $i_{\rm mut;0}=60^{\circ}, 80^{\circ}, 100^{\circ}$, and $120^{\circ}$, the mutual inclinations only change by $\Delta i_{\rm mut} = 4^{\circ}, 1^{\circ}, 1^{\circ}$, and $4^{\circ}$, respectively, for the eccentricity to reach the currently observed value. All of these values are consistent with the near-aligned stellar obliquity suggested by our combined fit measurement $\lambda =3.2^{+1.6}_{-1.5} \degree$.

One caveat in invoking the KL mechanism is that the torque exerted on the planet from the perturber in the KL mechanism is dependent on the pericenter argument of the planet ($\omega_{\rm p}$). On close-in orbits, precessions of the pericenter argument caused by other forces including general relativistic effects ($\left.\dot{\omega}_{\rm p}\right|_{\mathrm{GR}}$), rotational bulges ($\left.\dot{\omega}_{\rm p}\right|_{\text{rot}}$), and tidal deformations ($\left.\dot{\omega}_{\rm p}\right|_{\text{tide}}$) are strong and may suppress the KL oscillations. \citet{wu2003planet} summarizes the relative precession rates for general relativistic effects, rotational bulges, and tidal deformations as
\begin{equation}
    \begin{aligned}
    & \frac{\left.\dot{\omega}_{\rm p}\right|_{\mathrm{GR}}}{\left.\dot{\omega}_{\rm p}\right|_{\mathrm{Kozai}}} \approx 5900\left(\frac{a_{\rm p}}{0.47 \mathrm{au}}\right)^{-4}\left(\frac{a_c}{1000 \mathrm{au}}\right)^3 \\
    & \frac{\left.\dot{\omega}_{\rm p}\right|_{\text {rot}}}{\left.\dot{\omega}_{\rm p}\right|_{\text {Kozai }}} \approx 750\left(\frac{a_{\rm p}}{0.47 \mathrm{au}}\right)^{-8}\left(\frac{a_c}{1000 \mathrm{au}}\right)^3\left(\frac{R_{\rm p}}{\rj}\right)^5 \\
    & \frac{\left.\dot{\omega}_{\rm p}\right|_{\text {tide}}}{\left.\dot{\omega}_{\rm p}\right|_{\text {Kozai }}} \approx 270\left(\frac{a_{\rm p}}{0.47 \mathrm{au}}\right)^{-8}\left(\frac{a_c}{1000 \mathrm{au}}\right)^3\left(\frac{R_{\rm p}}{\rj}\right)^5,
    \end{aligned}
\end{equation}
where notations are the same as in Table \ref{tab:orbit}.
If any of these ratios exceeds unity, the KL oscillation is destroyed. 

For the current orbit of TOI-677 b as listed in Table \ref{tab:orbit}, the relative precession rates of the rotational bulges and tidal deformations both exceed unity, such that the planet's eccentricity cannot have been excited through the KL mechanism at its current location. However, it is possible that TOI-677 b's eccentricity was excited through the KL mechanism from an initial orbit further out in the system, and that the planet then migrated to its current position through tidal dissipation. During this migration process, the KL oscillation would be gradually quenched by higher-frequency precessions associated with general relativistic effects, rotational bulges, and tidal deformations. This is the same scenario as the high-eccentricity migration mechanism for hot Jupiters, except that the starting position is closer and the maximum eccentricity needed is lower \citep[e.g.][]{Wu_2023}. During tidal dissipation, the orbital angular momentum is conserved as \citep{dawson2018origins}
\begin{equation}\label{tide}
    a_{\rm p}(1-e_{\rm p}^2) = \text{constant}.
\end{equation}

For different starting positions, we use Equation \ref{tide} to estimate the maximum KL-excited eccentricity $e_{\rm KL}$ that must be reached to attain TOI-677 b's current observed orbital configuration. From this, we estimate the minimum initial mutual inclination $i_{\rm mut,0}$ required to excite the eccentricity from 0 to $e_{\rm KL}$ through the KL mechanism if the change in the mutual inclination is within 3-$\sigma$ of the measured $\lambda$ value ($\Delta i_{\rm mut} \lesssim 8 \degree$). We also inspect whether the eccentricity evolution timescale $\tau_{e [0.46]}$, during which the eccentricity changes from $e_{\rm KL}$ to the current observed value $0.46$, is shorter than the age of the system (about 3 Gyr from \citet{TOI677_Sedaghati_2023}). We calculated the eccentricity evolution timescale using Equation (6) to (13) in \citet{HJ_obliquity_Rice_2022}, adopting an effective tidal dissipation parameter $Q=10^5$ and Love number $k_2=0.3$ for TOI-677 b. Our calculated values for four sample initial orbital separations are listed in Table \ref{tab:tide}.

\begin{table}
    \centering
    \caption{Relative precession rates, maximum KL-excited eccentricity, closest distance, minimum initial mutual inclination, and eccentricity evolution timescales for four sample starting orbital separations of TOI-677 b.}
    \label{tab:tide}
    \begin{tabular}{lcccc}
    \hline\hline
    Parameter &  \multicolumn{4}{c}{Values}\\ 
    \hline
    $a_0$ (au) & 0.2 & 0.3 & 0.4 & 0.5 \\[0.1cm]
    $\Big(\frac{\left.\dot{\omega}_{\rm p}\right|_{\mathrm{GR}}}{\left.\dot{\omega}_{\rm p}\right|_{\mathrm{Kozai}}}\Big)$ &  0.0386 & 0.0076 & 0.0024 & 0.0010 \\[0.4cm]
    $\Big(\frac{\left.\dot{\omega}_{\rm p}\right|_{\text {rot }}}{\left.\dot{\omega}_{\rm p}\right|_{\text {Kozai}}}\Big)$ & 0.3283 & 0.0128 & 0.0013 & 0.0002 \\[0.4cm]
    $\Big(\frac{\left.\dot{\omega}_{\rm p}\right|_{\text {tide }}}{\left.\dot{\omega}_{\rm p}\right|_{\text {Kozai}}}\Big)$ & 0.1182 & 0.0046 & 0.0005 & 0.0001 \\[0.2cm]
    $e_{\rm KL}$ & 0.76 & 0.85 & 0.89 & 0.91 \\
    $a_0(1-e_{\rm KL})/R_\star$ & 7.79 & 7.43 & 7.28 & 7.19 \\
    $i_{\rm mut,0} \ (\degree)$ & 76.3 & 81.4 & 83.5 & 84.6 \\
    $\tau_{e [0.46]} \ (\rm Myr)$ & 34.4 & 15.5 & 9.2 & 6.2 \\
    \hline\hline
    \end{tabular}
\end{table}

If TOI-677 b had its eccentricity excited through the KL mechanism and then migrated to its current orbit through tidal dissipation, the angle between the planet's and companion's orbital planes ($i_{\rm mut}$) should fall within the range $70^{\degree}\lesssim i_{\rm mut}\lesssim110\degree$.

An important complication in this scenario is that, after the KL oscillation is quenched, the highly-inclined outer companion will cause the inner planet's orbit to undergo nodal precession, altering the inner planet's spin-orbit orientation and changing the true stellar obliquity \citep[e.g.][]{yee2018hatp11, bailey2020nodal}. The precession rate of the longitude of the ascending node of the inner planet ($\mathrm{d} \Omega_{\rm p} / \mathrm{d} t$) can be calculated as \citep[see Equation 3 in][]{yee2018hatp11}
\begin{equation}
        \frac{\mathrm{d} \Omega_{\rm p}}{\mathrm{d} t} = 
        \frac{n_{\rm p}}{2} \frac{M_{\rm c}}{M_\star} \left(\frac{a_{\rm p}}{a_{\rm c}\sqrt{1-e_{\rm c}^2}}\right)^3 \left(\frac{6+9e_{\rm p}^2}{\sqrt{1-e_{\rm p}^2}}\right)\cos i_{\rm mut},
\end{equation}
where $n_{\rm p}$ is the mean motion of the inner planet, and the meaning of other symbols is the same as above. The nodal precession period $\tau_{\rm nodal} \approx 2\pi / (\rm{d} \Omega_{\rm p} / \rm{d} t)$
of the current orbital configuration is approximately 0.14 Myr -- significantly shorter than the age of the system -- assuming a mutual inclination $i_{\rm mut} = 80\degree$. As a result, retaining a configuration with a low $\lambda$ would imply that the KL oscillation was quenched recently such that nodal precession has not yet changed the inner planet's orbital direction, or that the system is being observed at a low-stellar-obliquity trough in the precession cycle.  While not impossible, this need for a chance timing alignment casts doubt on the KL scenario for TOI-677 b.

\subsubsection{Eccentricity excitation through disk-planet interactions}\label{subsubsect:diske}
A third possible explanation for TOI-677 b's highly eccentric and aligned configuration is that the planet's eccentricity may have been excited through disk-planet interactions within an aligned, natal protoplanetary disk. However, most previous models of disk-planet interactions have struggled to excite planetary eccentricities at the level of $e\gtrsim0.3$ \citep[e.g.][]{kley2006disk, bitsch2010orbital, duffell2015eccentric}. 

Recently, several works proposed that the outer disk can excite the eccentricity of a planet in an inner cavity to higher values than previously attained. \citet{romanova2023higheindisk} demonstrated that, in the presence of Lindblad resonances between an outer disk and the planet in the inner cavity, the maximum possible eccentricity is pushed up to $e\sim0.65-0.75$. A caveat of this work is that the disk material cannot penetrate into the cavity, which excludes the local corotation torque that damps the planet's eccentricity. 

\citet{Li2023eccentricdisk} proposed that a planet within the inner cavity could, alternatively, attain a high eccentricity (between 0.1 and 0.6) through resonant interactions with a dispersing, low-eccentricity ($e\lesssim0.05$) protoplanetary disk. Although specific disk conditions, including the initial disk mass, gas temperature viscosity, and mass-loss rate, are needed for the disk to gain eccentricity through rapid cooling, this mechanism may operate in at least some regimes.

Furthermore, \citet{murray2022effects} demonstrated that if a resonant planet pair migrates into the cavity, the eccentricity of the inner planet can be excited to high values ($e>0.5$) through disk-driven differential precession between the pair. In this scenario, the disk must be heavy enough to produce a high differential precession rate between the two planets in resonance, and the disk dispersal must proceed slowly enough to avoid destabilizing the resonant planet pair. The outer companion's mass must be comparable to that of the inner planet for it to perturb the inner planet onto a high-eccentricity orbit, such that, if the resonant planet pair survived the disk dispersal, the companion's presence should have been inferred through either direct RV signatures or transit timing variations of TOI-677 b.

While eccentricity excitation through disk-planet interactions may be promising to produce eccentric and aligned warm-Jupiter systems, further work is needed to more clearly demonstrate whether the necessary initial conditions are typically met. For an individual planet TOI-677 b, it is difficult to definitively validate that its eccentricity is excited through disk-planet interactions. Nevertheless, this scenario would be strengthened if the outer brown dwarf companion is found to be coplanar, since a mutually inclined companion would be more indicative of dynamical evolution out of the disk plane.

\subsubsection{Implications for future observations}

The properties of TOI-677 b's outer companion offer important clues into potential formation scenarios for the system. The companion would most likely reside in a coplanar configuration if the eccentricity of TOI-677 b was excited through either disk-planet interactions or CHEM. 
Alternatively, the companion should have a substantial mutual orbital inclination with respect to the planet if TOI-677 b's high eccentricity was excited through the KL mechanism. 
Through further characterization of TOI-677 c -- from long-term RV observations and/or accurate astrometric characterization -- it may be possible to distinguish between the system's prospective formation channels and further clarify the origins of this enigmatic system. 

We anticipate that the ongoing \textit{Gaia} mission \citep{gaia2016} may provide useful constraints for TOI-677 c. 
The astrometric signature of TOI-677 c is $\alpha \sim 1.7\,\text{mas}$, about fifty times greater than the along-scan accuracy per field of view passage $\sigma_{\rm fov}=34.2\, \mu \text{as}$ for its $G = 9.661$ host star.
Therefore, according to the rough \textit{Gaia} detectability criterion $\alpha>2\sigma_{fov}$ in \citet{perryman2014gaia}, at least part of the orbital motion of TOI-677 c ($P_{\rm c}>10$ yr, \citet{TOI677_Sedaghati_2023}) should be detectable within the \textit{Gaia} mission baseline ($0.2-6 \ \text{yr}$).

\subsection{The alignment of warm Jupiters in single-star systems}\label{sect:alignment}

\begin{figure*}
    \centering
    \includegraphics[width=\textwidth]{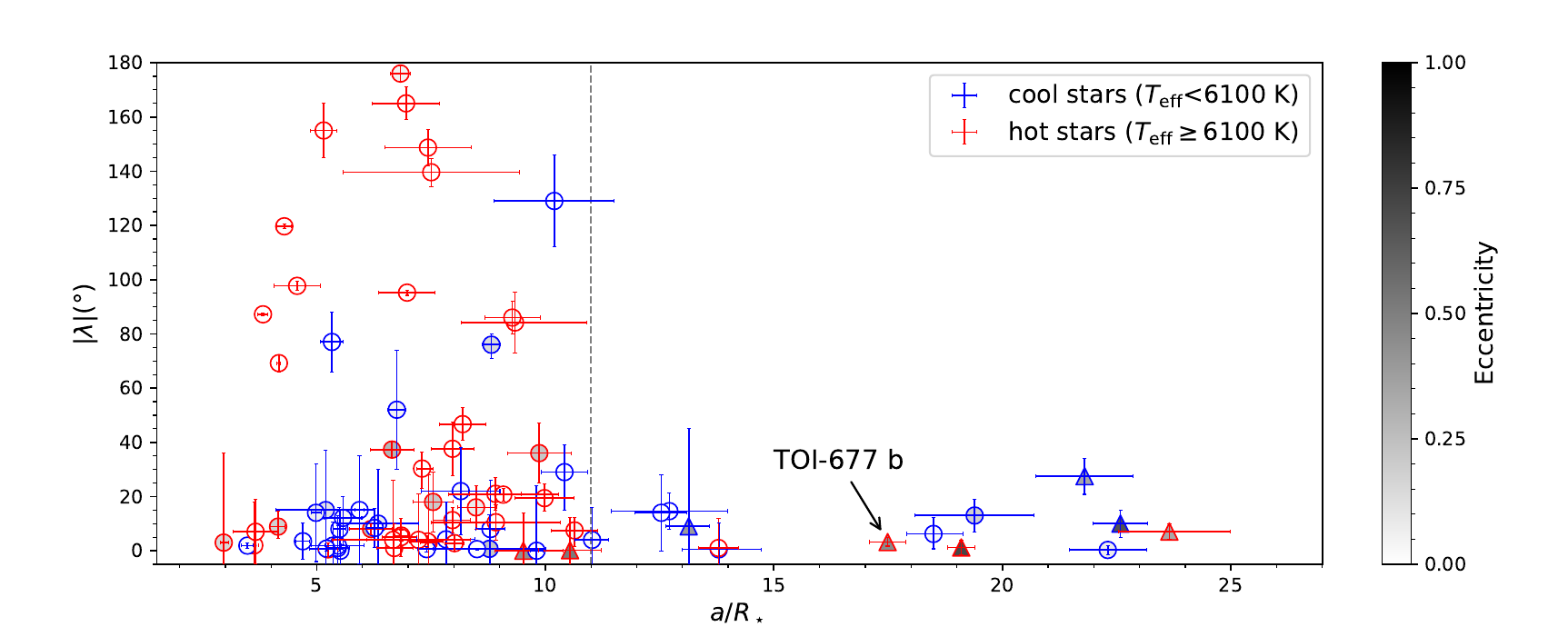}
    \caption{Distribution of measured $\left| \lambda \right|$ values for hot and warm Jupiters ($M_{\rm pl} \geq 0.4M_J$) in single-star systems as a function of the orbital separation $a/R_\star$. The vertical dashed gray line marks the rough boundary $a/R_\star=11$ between hot and warm Jupiters. The systems with hot stars ($\teff \geq 6100$ K) are shown as red-edged points, while those with cool stars ($\teff < 6100$ K) are shown as blue-edged points. The face colors of the points display the orbital eccentricity of the planet, and systems with $e>0.3$ are marked as triangles. The updated measurement of TOI-677 b from this work is labeled in the figure.}
    \label{fig:lambdas}
\end{figure*}

 \citet{rice2022tendency} examined the sky-projected stellar obliquity distribution in relatively isolated, single-star systems and found that, in these systems, tidally detached Jovian planets are preferentially more aligned than their closer-orbiting counterparts. Over the 20 months since that sample was developed, several more stellar obliquity measurements of tidally-detached warm Jupiter systems have been accumulated. Here, we revisit the sky-projected stellar obliquity distribution\footnote{While the sky-projected spin-orbit angle $\lambda$ is not equivalent to the true spin-orbit angle $\psi$, the 3-D stellar obliquity $\psi$ is related to and always closer to $90\degree$ than the 2-D stellar obliquity $\lambda$ \citep{fabrycky2009psi}. \citet{dong2023hierarchicalD} also demonstrated that, at the population level, the distribution of $\lambda$ is roughly equivalent to the underlying $\psi$ distribution. Therefore, we can use the sky-projected spin-orbit angle $\lambda$ to study the statistical properties of the stellar obliquity in close-in gas giant systems.} of hot and warm Jupiters in single-star systems with these newer measurements incorporated. 

 Our sample consists of systems included in both the \citet{ps_nasaexo} ``Planetary Systems'' list and the TEPCat catalog\footnote{\url{https://www.astro.keele.ac.uk/jkt/tepcat/}} \citep{southworth2011homogeneous}, which compiles published stellar obliquity measurements into a single database. Both datasets were downloaded on 2023 October 20. We drew $\lambda$ and $\teff$ from the TEPCat catalog, and all other system parameters were drawn from the default parameter set of the NASA Exoplanet Archive. We used the most recently obtained stellar obliquity measurement for most stars, with the exception of systems in which previous measurements were much more precise. Exoplanets with no eccentricity measurements in the NASA Exoplanet Archive were excluded from the sample.
 
From this combined dataset, we selected gas giants in single-star systems for our analysis. Following \citet{rice2022tendency}, we used $M_{\rm pl} \geq 0.4 \mj$ as the cut-off for Jovian planets, $a/R_\star=11$ as the dividing line between close-orbiting hot Jupiters and tidally-detached warm Jupiters, and $\teff = 6100$ K (the Kraft break) as the boundary between hot and cool stars. Single-star systems were selected by removing any systems with \texttt{sy\_snum > 1} in the NASA Exoplanet Archive, as well as any systems identified with comoving companions from the \textit{Gaia} DR3 catalogue \citep{brown2023gaiadr3} following the methods of \citet{el2021million} and \citet{rice2022tendency}. We note that this selection explicitly removes ``Saturns'' ($0.2 \mj \lesssim M_{\rm pl} \lesssim 0.4 \mj$) -- the lowest-mass Jovian planets that previous samples have shown are more often misaligned than higher-mass Jovians, and that do not appear to follow the same trends \citep{rice2022tendency, morgan2023signs}.

Figure \ref{fig:lambdas} places TOI-677 b into context as one of the widest-separation tidally detached exoplanets with a measured $\lambda$ value. All of the warm Jupiters in single-star systems are, so far, consistent with alignment ($\left| \lambda \right|\leq20\degree$) within 1-$\sigma$ \citep{harre2023wasp106b, espinozaretamal2023aligned}. Despite their small spin-orbit angles, the eccentricity distribution of warm Jupiters spans a wide range of values, with an even broader dispersion than that of their close-orbiting counterparts \citep{yu2018epic}.

Moreover, TOI-677 b supports the previously suggested, tentative trend that wide-orbiting Jupiters around hot stars may be preferentially more aligned than those around cool stars \citep{rice2021soles}. If this trend continues to gain significance, it may be indicative of an obliquity excitation mechanism restricted to cool stars. For example, \citet{anderson2018teetering} proposed that, in the presence of an external, inclined Jovian-mass companion, spin–orbit misalignments can be excited preferentially in cool-star systems due to a secular resonance between the host star's spin axis precession frequency and nodal precession induced by interactions with the companion. Further monitoring to constrain the companion rate of misaligned warm Jupiters around cool stars would be necessary to confirm this hypothesis.

\subsection{Toward an era of extreme-precision spin-orbit characterization}\label{subsection:diff}

Existing Rossiter-McLaughlin measurements of wide-orbiting giant planets have shown that warm Jupiter exoplanets may follow an intrinsically different spin-orbit angle distribution from that of their close-orbiting counterparts \citep{rice2021soles, rice2022tendency}. However, the true distribution of spin-orbit angles for each population remains elusive due to a combination of small-number statistics and often-large uncertainties for individual measurements. Even those systems that are thought to form quiescently and would generally be classified as ``aligned'' demonstrate some spread in their measured spin-orbit angles \citep{rice2023evidence}. 

Figure \ref{fig:RM} demonstrates that the main factor driving differences between the PFS model and the ESPRESSO model for TOI-677 b is the highest peak of the ESPRESSO dataset, where two points deviate from the rest of the trend and consequently push $\lambda$ toward smaller values. These outliers may have originated from stellar-activity-induced jitter, demonstrating the benefit of repeat measurements when deriving fundamental parameters at high precision. In particular, our work demonstrates the importance of repeat measurements when constraining the \textit{degree} of alignment of a given planetary system.

One potential alternative is that the stellar obliquity of the system may vary slightly as a function of time. This possibility has been examined in the past for the XO-3 system, where multiple measurements were discrepant by tens of degrees ($70\pm15\degree$ in \citet{hebrard2008misaligned}; $37.3\pm3.7\degree$ in \citet{winn2009on}; $37.3\pm3.0\degree$ in \citet{hirano2011further}); however, \citet{worku2022revisiting} found that there was no strong evidence for a change over time in the spin-orbit angle of XO-3 b. If present, temporal changes in the spin-orbit orientation may be caused by angular momentum transport from internal gravity waves within the host star \citep{rogers2012internal}.

The origins of true low-level misalignments -- which, as demonstrated by this work, must be disentangled from signatures of stellar activity -- may offer insights into our own solar system's $\psi=6\degree$ stellar obliquity \citep{souami2012solar}, which lies near, but is decidedly not in exact, alignment. Analyses such as the one presented here, with more than one in-transit dataset incorporated, help to push beyond the broad population-wide regimes of ``aligned'' versus ``misaligned'' exoplanets and toward extreme-precision constraints that may demonstrate the typical degree of spin-orbit alignment across different types of planetary systems \citep{albrecht2022stellar}.

\section{Conclusions}\label{sect:conclusion}
We have presented a new Rossiter-McLaughlin measurement of the tidally-detached warm Jupiter TOI-677 b from PFS/Magellan. We combined this new measurement with archival data, including a recently published Rossiter-McLaughlin measurement from the ESPRESSO spectrograph \citep{TOI677_Sedaghati_2023}, to derive a self-consistent set of updated physical parameters for TOI-677 b. Our independent observation corroborates previous findings from \citet{TOI677_Sedaghati_2023} that TOI-677 b exhibits a high eccentricity and yet a surprisingly low sky-projected spin-orbit angle, with values $e=0.460^{+0.019}_{-0.018}$ and $\lambda=3.2^{+1.6}_{-1.5}\degree$ from our combined fit.

Several formation pathways for TOI-677 b were examined through this work, each aiming to preserve the planet's low observed sky-projected stellar obliquity while reproducing its high eccentricity. Ultimately, we find that CHEM is disfavored by the measured properties of the outer brown dwarf companion, and KL oscillations followed by tidal dissipation, while feasible, would require a low-likelihood chance alignment due to nodal precession from the outer companion. As a result, we conclude hat TOI-677 b most likely had its eccentricity excited through disk-planet interactions. Further monitoring of the outer companion -- in particular, improved constraints on its mass and mutual inclination relative to TOI-677 b -- would help to confirm this conjecture. 

Our results expand upon previous findings from \citet{rice2022tendency}, showing that warm Jupiters in single-star systems continue to exhibit minimal deviation from spin-orbit alignment despite an emerging dispersion in orbital eccentricity and host star type.
Ultimately, a larger population of warm Jupiters with precise spin-orbit measurements, spanning a wider range of orbital eccentricities and host star types, will be necessary to more fully demonstrate the range of orbital configurations spanned by warm Jupiters and the corresponding implications for their formation.

\section{Acknowledgments}
\label{section:acknowledgments}

Q.H. is very grateful to Zhang Yisong at the Physics Department of Tsinghua University for his support in the discussion concerning the Kozai-Lidov oscillations. Q.H. thanks Prof. Zhu Wei from the Astronomy Department of Tsinghua University for his advice on the discussion sections. M.R. acknowledges support from Heising-Simons Foundation Grants \#2023-4478 and \#2023-4655, NASA grant GR122985/AWD0011072, and Oracle for Research grant No. CPQ-3033929. S.W. acknowledges support from Heising-Simons Foundation Grant \#2023-4050.

This work has benefited from the use of the \textit{Grace} computing cluster at the Yale Center for Research Computing (YCRC).
This work makes use of observations from the Las Cumbres Observatory global telescope (LCOGT) network. Part of the LCOGT telescope time was granted by NOIRLab through the Mid-Scale Innovations Program (MSIP). MSIP is funded by NSF.
This paper is based on observations made with the Las Cumbres Observatory’s education network telescopes that were upgraded through generous support from the Gordon and Betty Moore Foundation.
This research has made use of the Exoplanet Follow-up Observation Program (ExoFOP; DOI: \dataset[10.26134/ExoFOP5]{http://dx.doi.org/10.26134/ExoFOP5}) website, which is operated by the California Institute of Technology, under contract with the National Aeronautics and Space Administration under the Exoplanet Exploration Program.
This research has made use of the \citet{ps_nasaexo} (DOI: \dataset[10.26133/NEA12]{http://dx.doi.org/10.26133/NEA12}), which is operated by the California Institute of Technology, under contract with the National Aeronautics and Space Administration under the Exoplanet Exploration Program.
This paper includes data collected by the TESS mission \citep{tess_all_sectors} (DOI: \dataset[10.17909/t9-nmc8-f686]{http://dx.doi.org/10.17909/t9-nmc8-f686}). Funding for the TESS mission is provided by NASA's Science Mission Directorate. K.A.C. acknowledges support from the TESS mission via subaward s3449 from MIT.

\software{\texttt{allesfitter} \citep{gunther2021allesfitter}, \texttt{emcee} \citep{foremanmackey2013}, \texttt{ellc} \citep{maxted2016ellc}, \texttt{lightkurve} \citep{cardoso2018lightkurve}, \texttt{juliet} \citep{Espinoza2019_Juliet}, \texttt{matplotlib} \citep{hunter2007matplotlib}, \texttt{numpy} \citep{oliphant2006guide, walt2011numpy, harris2020array}, \texttt{pandas} \citep{mckinney2010data}, \texttt{scipy} \citep{virtanen2020scipy}, AstroImageJ \citep{Collins:2017}, TAPIR \citep{Jensen:2013}}

\facility{Magellan II: PFS, LCOGT CTIO 1m ($\times2$) and 0.4m ($\times2$), TEPCat}

\bibliography{bibliography}

\begin{thebibliography}{}
\expandafter\ifx\csname natexlab\endcsname\relax\def\natexlab#1{#1}\fi
\providecommand{\url}[1]{\href{#1}{#1}}

\bibitem[{{Addison} {et~al.}(2018){Addison}, {Wang}, {Johnson}, {Tinney},
  {Wright}, \& {Bayliss}}]{addison2018polar}
{Addison}, B.~C., {Wang}, S., {Johnson}, M.~C., {et~al.} 2018, \aj, 156, 197

\bibitem[{Albrecht {et~al.}(2012)Albrecht, Winn, Johnson, Howard, Marcy,
  Butler, Arriagada, Crane, Shectman, Thompson,
  {et~al.}}]{albrecht2012obliquities}
Albrecht, S., Winn, J.~N., Johnson, J.~A., {et~al.} 2012, \apj, 757, 18

\bibitem[{{Albrecht} {et~al.}(2022){Albrecht}, {Dawson}, \&
  {Winn}}]{albrecht2022stellar}
{Albrecht}, S.~H., {Dawson}, R.~I., \& {Winn}, J.~N. 2022, \pasp, 134, 082001

\bibitem[{Albrecht {et~al.}(2021)Albrecht, Marcussen, Winn, Dawson, \&
  Knudstrup}]{albrecht2021preponderance}
Albrecht, S.~H., Marcussen, M.~L., Winn, J.~N., Dawson, R.~I., \& Knudstrup, E.
  2021, \apjl, 916, L1

\bibitem[{{Anderson} {et~al.}(2018){Anderson}, {Temple}, {Nielsen}, {Burdanov},
  {Hellier}, {Bouchy}, {Brown}, {Collier Cameron}, {Gillon}, {Jehin}, {Maxted},
  {Pepe}, {Pollacco}, {Pozuelos}, {Queloz}, {S{\'e}gransan}, {Smalley},
  {Triaud}, {Turner}, {Udry}, \& {West}}]{anderson2018wasp}
{Anderson}, D.~R., {Temple}, L.~Y., {Nielsen}, L.~D., {et~al.} 2018, arXiv
  e-prints, arXiv:1809.04897

\bibitem[{Anderson \& Lai(2018)}]{anderson2018teetering}
Anderson, K.~R., \& Lai, D. 2018, \mnras, 480, 1402

\bibitem[{{Anderson} {et~al.}(2016){Anderson}, {Storch}, \&
  {Lai}}]{andreson2016kl}
{Anderson}, K.~R., {Storch}, N.~I., \& {Lai}, D. 2016, \mnras, 456, 3671

\bibitem[{{Attia} {et~al.}(2023){Attia}, {Bourrier}, {Delisle}, \&
  {Eggenberger}}]{attia2023DREAM}
{Attia}, O., {Bourrier}, V., {Delisle}, J.~B., \& {Eggenberger}, P. 2023, \aap,
  674, A120

\bibitem[{Bailey \& Fabrycky(2020)}]{bailey2020nodal}
Bailey, N., \& Fabrycky, D. 2020, The Astronomical Journal, 159, 217.
\newblock \url{https://dx.doi.org/10.3847/1538-3881/ab83f0}

\bibitem[{Beaugé \& Nesvorný(2012)}]{beauge2012planetscatter}
Beaugé, C., \& Nesvorný, D. 2012, The Astrophysical Journal, 751, 119.
\newblock \url{https://dx.doi.org/10.1088/0004-637X/751/2/119}

\bibitem[{{Bitsch} \& {Kley}(2010)}]{bitsch2010orbital}
{Bitsch}, B., \& {Kley}, W. 2010, \aap, 523, A30

\bibitem[{{Bourrier} {et~al.}(2020){Bourrier}, {Ehrenreich}, {Lendl},
  {Cretignier}, {Allart}, {Dumusque}, {Cegla}, {Su{\'a}rez-Mascare{\~n}o},
  {Wyttenbach}, {Hoeijmakers}, {Melo}, {Kuntzer}, {Astudillo-Defru}, {Giles},
  {Heng}, {Kitzmann}, {Lavie}, {Lovis}, {Murgas}, {Nascimbeni}, {Pepe}, {Pino},
  {Segransan}, \& {Udry}}]{bourrier2020hot}
{Bourrier}, V., {Ehrenreich}, D., {Lendl}, M., {et~al.} 2020, \aap, 635, A205

\bibitem[{{Brown} {et~al.}(2013){Brown}, {Baliber}, {Bianco}, {Bowman},
  {Burleson}, {Conway}, {Crellin}, {Depagne}, {De Vera}, {Dilday}, {Dragomir},
  {Dubberley}, {Eastman}, {Elphick}, {Falarski}, {Foale}, {Ford}, {Fulton},
  {Garza}, {Gomez}, {Graham}, {Greene}, {Haldeman}, {Hawkins}, {Haworth},
  {Haynes}, {Hidas}, {Hjelstrom}, {Howell}, {Hygelund}, {Lister}, {Lobdill},
  {Martinez}, {Mullins}, {Norbury}, {Parrent}, {Paulson}, {Petry}, {Pickles},
  {Posner}, {Rosing}, {Ross}, {Sand}, {Saunders}, {Shobbrook}, {Shporer},
  {Street}, {Thomas}, {Tsapras}, {Tufts}, {Valenti}, {Vander Horst}, {Walker},
  {White}, \& {Willis}}]{Brown:2013}
{Brown}, T.~M., {Baliber}, N., {Bianco}, F.~B., {et~al.} 2013, \pasp, 125, 1031

\bibitem[{Butler {et~al.}(1996)Butler, Marcy, Williams, McCarthy, Dosanjh, \&
  Vogt}]{butler1996attaining}
Butler, R.~P., Marcy, G.~W., Williams, E., {et~al.} 1996, Publications of the
  Astronomical Society of the Pacific, 108, 500

\bibitem[{Cardoso {et~al.}(2018)Cardoso, Hedges, Gully-Santiago, Saunders,
  Cody, Barclay, Hall, Sagear, Turtelboom, Zhang,
  {et~al.}}]{cardoso2018lightkurve}
Cardoso, J. V. d.~M., Hedges, C., Gully-Santiago, M., {et~al.} 2018,
  Astrophysics Source Code Library, ascl

\bibitem[{{Collins} {et~al.}(2017){Collins}, {Kielkopf}, {Stassun}, \&
  {Hessman}}]{Collins:2017}
{Collins}, K.~A., {Kielkopf}, J.~F., {Stassun}, K.~G., \& {Hessman}, F.~V.
  2017, \aj, 153, 77

\bibitem[{Crane {et~al.}(2006)Crane, Shectman, \& Butler}]{crane2006carnegie}
Crane, J.~D., Shectman, S.~A., \& Butler, R.~P. 2006, in Ground-based and
  Airborne Instrumentation for Astronomy, Vol. 6269, SPIE, 972--982

\bibitem[{Crane {et~al.}(2010)Crane, Shectman, Butler, Thompson, Birk, Jones,
  \& Burley}]{crane2010carnegie}
Crane, J.~D., Shectman, S.~A., Butler, R.~P., {et~al.} 2010, in Ground-based
  and Airborne Instrumentation for Astronomy III, Vol. 7735, SPIE, 1909--1923

\bibitem[{Crane {et~al.}(2008)Crane, Shectman, Butler, Thompson, \&
  Burley}]{crane2008carnegie}
Crane, J.~D., Shectman, S.~A., Butler, R.~P., Thompson, I.~B., \& Burley, G.~S.
  2008, in Ground-based and Airborne Instrumentation for Astronomy II, Vol.
  7014, SPIE, 2484--2493

\bibitem[{Dawson \& Johnson(2018)}]{dawson2018origins}
Dawson, R.~I., \& Johnson, J.~A. 2018, \araa, 56, 175

\bibitem[{{Dong} \& {Foreman-Mackey}(2023)}]{dong2023hierarchicalD}
{Dong}, J., \& {Foreman-Mackey}, D. 2023, \aj, 166, 112

\bibitem[{{Dong} {et~al.}(2023){Dong}, {Wang}, {Rice}, {Zhou}, {Huang},
  {Dawson}, {Stef{\'a}nsson}, {Halverson}, {Kanodia}, {Mahadevan}, {McElwain},
  {Alvarado-Montes}, {Ninan}, {Robertson}, {Roy}, {Schwab}, {Logsdon},
  {Terrien}, {Collins}, {Srdoc}, {Sefako}, {Laloum}, {Latham}, {Bieryla},
  {Dalba}, {Dragomir}, {Villanueva}, {Howell}, {Ricker}, {Seager}, {Winn},
  {Jenkins}, {Shporer}, \& {Rapetti}}]{dong2023toi}
{Dong}, J., {Wang}, S., {Rice}, M., {et~al.} 2023, \apjl, 951, L29

\bibitem[{{Duffell} \& {Chiang}(2015)}]{duffell2015eccentric}
{Duffell}, P.~C., \& {Chiang}, E. 2015, \apj, 812, 94

\bibitem[{El-Badry {et~al.}(2021)El-Badry, Rix, \& Heintz}]{el2021million}
El-Badry, K., Rix, H.-W., \& Heintz, T.~M. 2021, \mnras, 506, 2269

\bibitem[{{Espinoza} {et~al.}(2019){Espinoza}, {Kossakowski}, \&
  {Brahm}}]{Espinoza2019_Juliet}
{Espinoza}, N., {Kossakowski}, D., \& {Brahm}, R. 2019, \mnras, 490, 2262

\bibitem[{Espinoza-Retamal {et~al.}(2023)Espinoza-Retamal, Brahm, Petrovich,
  Jordán, Stefánsson, Sedaghati, Hobson, Muñoz, Boyle, Leiva, \&
  Suc}]{espinozaretamal2023aligned}
Espinoza-Retamal, J.~I., Brahm, R., Petrovich, C., {et~al.} 2023, The Aligned
  Orbit of the Eccentric Proto Hot Jupiter TOI-3362b, , , arXiv:2309.03306

\bibitem[{Fabrycky \& Tremaine(2007)}]{fabrycky2007shrinking}
Fabrycky, D., \& Tremaine, S. 2007, \apj, 669, 1298

\bibitem[{Fabrycky \& Winn(2009)}]{fabrycky2009psi}
Fabrycky, D.~C., \& Winn, J.~N. 2009, The Astrophysical Journal, 696, 1230.
\newblock \url{https://dx.doi.org/10.1088/0004-637X/696/2/1230}

\bibitem[{{Foreman-Mackey} {et~al.}(2017){Foreman-Mackey}, {Agol}, {Angus}, \&
  {Ambikasaran}}]{Foreman-Mackey2017_celerite}
{Foreman-Mackey}, D., {Agol}, E., {Angus}, R., \& {Ambikasaran}, S. 2017, AJ,
  154, 220.
\newblock \url{https://arxiv.org/abs/1703.09710}

\bibitem[{{Foreman-Mackey} {et~al.}(2013{\natexlab{a}}){Foreman-Mackey},
  {Hogg}, {Lang}, \& {Goodman}}]{emcee}
{Foreman-Mackey}, D., {Hogg}, D.~W., {Lang}, D., \& {Goodman}, J.
  2013{\natexlab{a}}, \pasp, 125, 306

\bibitem[{{Foreman-Mackey} {et~al.}(2013{\natexlab{b}}){Foreman-Mackey},
  {Hogg}, {Lang}, \& {Goodman}}]{foremanmackey2013}
---. 2013{\natexlab{b}}, \pasp, 125, 306

\bibitem[{{Gaia Collaboration} {et~al.}(2016){Gaia Collaboration}, {Prusti},
  {de Bruijne}, {Brown}, {Vallenari}, {Babusiaux}, {Bailer-Jones}, {Bastian},
  {Biermann}, {Evans}, {Eyer}, {Jansen}, {Jordi}, {Klioner}, {Lammers},
  {Lindegren}, {Luri}, {Mignard}, {Milligan}, {Panem}, {Poinsignon},
  {Pourbaix}, {Randich}, {Sarri}, {Sartoretti}, {Siddiqui}, {Soubiran},
  {Valette}, {van Leeuwen}, {Walton}, {Aerts}, {Arenou}, {Cropper}, {Drimmel},
  {H{\o}g}, {Katz}, {Lattanzi}, {O'Mullane}, {Grebel}, {Holland}, {Huc},
  {Passot}, {Bramante}, {Cacciari}, {Casta{\~n}eda}, {Chaoul}, {Cheek}, {De
  Angeli}, {Fabricius}, {Guerra}, {Hern{\'a}ndez}, {Jean-Antoine-Piccolo},
  {Masana}, {Messineo}, {Mowlavi}, {Nienartowicz}, {Ord{\'o}{\~n}ez-Blanco},
  {Panuzzo}, {Portell}, {Richards}, {Riello}, {Seabroke}, {Tanga},
  {Th{\'e}venin}, {Torra}, {Els}, {Gracia-Abril}, {Comoretto},
  {Garcia-Reinaldos}, {Lock}, {Mercier}, {Altmann}, {Andrae}, {Astraatmadja},
  {Bellas-Velidis}, {Benson}, {Berthier}, {Blomme}, {Busso}, {Carry},
  {Cellino}, {Clementini}, {Cowell}, {Creevey}, {Cuypers}, {Davidson}, {De
  Ridder}, {de Torres}, {Delchambre}, {Dell'Oro}, {Ducourant}, {Fr{\'e}mat},
  {Garc{\'\i}a-Torres}, {Gosset}, {Halbwachs}, {Hambly}, {Harrison}, {Hauser},
  {Hestroffer}, {Hodgkin}, {Huckle}, {Hutton}, {Jasniewicz}, {Jordan},
  {Kontizas}, {Korn}, {Lanzafame}, {Manteiga}, {Moitinho}, {Muinonen},
  {Osinde}, {Pancino}, {Pauwels}, {Petit}, {Recio-Blanco}, {Robin}, {Sarro},
  {Siopis}, {Smith}, {Smith}, {Sozzetti}, {Thuillot}, {van Reeven}, {Viala},
  {Abbas}, {Abreu Aramburu}, {Accart}, {Aguado}, {Allan}, {Allasia},
  {Altavilla}, {{\'A}lvarez}, {Alves}, {Anderson}, {Andrei}, {Anglada Varela},
  {Antiche}, {Antoja}, {Ant{\'o}n}, {Arcay}, {Atzei}, {Ayache}, {Bach},
  {Baker}, {Balaguer-N{\'u}{\~n}ez}, {Barache}, {Barata}, {Barbier}, {Barblan},
  {Baroni}, {Barrado y Navascu{\'e}s}, {Barros}, {Barstow}, {Becciani},
  {Bellazzini}, {Bellei}, {Bello Garc{\'\i}a}, {Belokurov}, {Bendjoya},
  {Berihuete}, {Bianchi}, {Bienaym{\'e}}, {Billebaud}, {Blagorodnova},
  {Blanco-Cuaresma}, {Boch}, {Bombrun}, {Borrachero}, {Bouquillon}, {Bourda},
  {Bouy}, {Bragaglia}, {Breddels}, {Brouillet}, {Br{\"u}semeister},
  {Bucciarelli}, {Budnik}, {Burgess}, {Burgon}, {Burlacu}, {Busonero}, {Buzzi},
  {Caffau}, {Cambras}, {Campbell}, {Cancelliere}, {Cantat-Gaudin}, {Carlucci},
  {Carrasco}, {Castellani}, {Charlot}, {Charnas}, {Charvet}, {Chassat},
  {Chiavassa}, {Clotet}, {Cocozza}, {Collins}, {Collins}, {Costigan}, {Crifo},
  {Cross}, {Crosta}, {Crowley}, {Dafonte}, {Damerdji}, {Dapergolas}, {David},
  {David}, {De Cat}, {de Felice}, {de Laverny}, {De Luise}, {De March}, {de
  Martino}, {de Souza}, {Debosscher}, {del Pozo}, {Delbo}, {Delgado},
  {Delgado}, {di Marco}, {Di Matteo}, {Diakite}, {Distefano}, {Dolding}, {Dos
  Anjos}, {Drazinos}, {Dur{\'a}n}, {Dzigan}, {Ecale}, {Edvardsson}, {Enke},
  {Erdmann}, {Escolar}, {Espina}, {Evans}, {Eynard Bontemps}, {Fabre},
  {Fabrizio}, {Faigler}, {Falc{\~a}o}, {Farr{\`a}s Casas}, {Faye}, {Federici},
  {Fedorets}, {Fern{\'a}ndez-Hern{\'a}ndez}, {Fernique}, {Fienga}, {Figueras},
  {Filippi}, {Findeisen}, {Fonti}, {Fouesneau}, {Fraile}, {Fraser}, {Fuchs},
  {Furnell}, {Gai}, {Galleti}, {Galluccio}, {Garabato}, {Garc{\'\i}a-Sedano},
  {Gar{\'e}}, {Garofalo}, {Garralda}, {Gavras}, {Gerssen}, {Geyer}, {Gilmore},
  {Girona}, {Giuffrida}, {Gomes}, {Gonz{\'a}lez-Marcos},
  {Gonz{\'a}lez-N{\'u}{\~n}ez}, {Gonz{\'a}lez-Vidal}, {Granvik}, {Guerrier},
  {Guillout}, {Guiraud}, {G{\'u}rpide}, {Guti{\'e}rrez-S{\'a}nchez}, {Guy},
  {Haigron}, {Hatzidimitriou}, {Haywood}, {Heiter}, {Helmi}, {Hobbs},
  {Hofmann}, {Holl}, {Holland}, {Hunt}, {Hypki}, {Icardi}, {Irwin}, {Jevardat
  de Fombelle}, {Jofr{\'e}}, {Jonker}, {Jorissen}, {Julbe}, {Karampelas},
  {Kochoska}, {Kohley}, {Kolenberg}, {Kontizas}, {Koposov}, {Kordopatis},
  {Koubsky}, {Kowalczyk}, {Krone-Martins}, {Kudryashova}, {Kull}, {Bachchan},
  {Lacoste-Seris}, {Lanza}, {Lavigne}, {Le Poncin-Lafitte}, {Lebreton},
  {Lebzelter}, {Leccia}, {Leclerc}, {Lecoeur-Taibi}, {Lemaitre}, {Lenhardt},
  {Leroux}, {Liao}, {Licata}, {Lindstr{\o}m}, {Lister}, {Livanou}, {Lobel},
  {L{\"o}ffler}, {L{\'o}pez}, {Lopez-Lozano}, {Lorenz}, {Loureiro},
  {MacDonald}, {Magalh{\~a}es Fernandes}, {Managau}, {Mann}, {Mantelet},
  {Marchal}, {Marchant}, {Marconi}, {Marie}, {Marinoni}, {Marrese},
  {Marschalk{\'o}}, {Marshall}, {Mart{\'\i}n-Fleitas}, {Martino}, {Mary},
  {Matijevi{\v{c}}}, {Mazeh}, {McMillan}, {Messina}, {Mestre}, {Michalik},
  {Millar}, {Miranda}, {Molina}, {Molinaro}, {Molinaro}, {Moln{\'a}r},
  {Moniez}, {Montegriffo}, {Monteiro}, {Mor}, {Mora}, {Morbidelli}, {Morel},
  {Morgenthaler}, {Morley}, {Morris}, {Mulone}, {Muraveva}, {Musella},
  {Narbonne}, {Nelemans}, {Nicastro}, {Noval}, {Ord{\'e}novic},
  {Ordieres-Mer{\'e}}, {Osborne}, {Pagani}, {Pagano}, {Pailler}, {Palacin},
  {Palaversa}, {Parsons}, {Paulsen}, {Pecoraro}, {Pedrosa}, {Pentik{\"a}inen},
  {Pereira}, {Pichon}, {Piersimoni}, {Pineau}, {Plachy}, {Plum}, {Poujoulet},
  {Pr{\v{s}}a}, {Pulone}, {Ragaini}, {Rago}, {Rambaux}, {Ramos-Lerate},
  {Ranalli}, {Rauw}, {Read}, {Regibo}, {Renk}, {Reyl{\'e}}, {Ribeiro},
  {Rimoldini}, {Ripepi}, {Riva}, {Rixon}, {Roelens}, {Romero-G{\'o}mez},
  {Rowell}, {Royer}, {Rudolph}, {Ruiz-Dern}, {Sadowski}, {Sagrist{\`a}
  Sell{\'e}s}, {Sahlmann}, {Salgado}, {Salguero}, {Sarasso}, {Savietto},
  {Schnorhk}, {Schultheis}, {Sciacca}, {Segol}, {Segovia}, {Segransan},
  {Serpell}, {Shih}, {Smareglia}, {Smart}, {Smith}, {Solano}, {Solitro},
  {Sordo}, {Soria Nieto}, {Souchay}, {Spagna}, {Spoto}, {Stampa}, {Steele},
  {Steidelm{\"u}ller}, {Stephenson}, {Stoev}, {Suess}, {S{\"u}veges}, {Surdej},
  {Szabados}, {Szegedi-Elek}, {Tapiador}, {Taris}, {Tauran}, {Taylor},
  {Teixeira}, {Terrett}, {Tingley}, {Trager}, {Turon}, {Ulla}, {Utrilla},
  {Valentini}, {van Elteren}, {Van Hemelryck}, {van Leeuwen}, {Varadi},
  {Vecchiato}, {Veljanoski}, {Via}, {Vicente}, {Vogt}, {Voss}, {Votruba},
  {Voutsinas}, {Walmsley}, {Weiler}, {Weingrill}, {Werner}, {Wevers},
  {Whitehead}, {Wyrzykowski}, {Yoldas}, {{\v{Z}}erjal}, {Zucker}, {Zurbach},
  {Zwitter}, {Alecu}, {Allen}, {Allende Prieto}, {Amorim},
  {Anglada-Escud{\'e}}, {Arsenijevic}, {Azaz}, {Balm}, {Beck}, {Bernstein},
  {Bigot}, {Bijaoui}, {Blasco}, {Bonfigli}, {Bono}, {Boudreault}, {Bressan},
  {Brown}, {Brunet}, {Bunclark}, {Buonanno}, {Butkevich}, {Carret}, {Carrion},
  {Chemin}, {Ch{\'e}reau}, {Corcione}, {Darmigny}, {de Boer}, {de Teodoro}, {de
  Zeeuw}, {Delle Luche}, {Domingues}, {Dubath}, {Fodor}, {Fr{\'e}zouls},
  {Fries}, {Fustes}, {Fyfe}, {Gallardo}, {Gallegos}, {Gardiol}, {Gebran},
  {Gomboc}, {G{\'o}mez}, {Grux}, {Gueguen}, {Heyrovsky}, {Hoar}, {Iannicola},
  {Isasi Parache}, {Janotto}, {Joliet}, {Jonckheere}, {Keil}, {Kim},
  {Klagyivik}, {Klar}, {Knude}, {Kochukhov}, {Kolka}, {Kos}, {Kutka}, {Lainey},
  {LeBouquin}, {Liu}, {Loreggia}, {Makarov}, {Marseille}, {Martayan},
  {Martinez-Rubi}, {Massart}, {Meynadier}, {Mignot}, {Munari}, {Nguyen},
  {Nordlander}, {Ocvirk}, {O'Flaherty}, {Olias Sanz}, {Ortiz}, {Osorio},
  {Oszkiewicz}, {Ouzounis}, {Palmer}, {Park}, {Pasquato}, {Peltzer}, {Peralta},
  {P{\'e}turaud}, {Pieniluoma}, {Pigozzi}, {Poels}, {Prat}, {Prod'homme},
  {Raison}, {Rebordao}, {Risquez}, {Rocca-Volmerange}, {Rosen}, {Ruiz-Fuertes},
  {Russo}, {Sembay}, {Serraller Vizcaino}, {Short}, {Siebert}, {Silva},
  {Sinachopoulos}, {Slezak}, {Soffel}, {Sosnowska}, {Strai{\v{z}}ys}, {ter
  Linden}, {Terrell}, {Theil}, {Tiede}, {Troisi}, {Tsalmantza}, {Tur},
  {Vaccari}, {Vachier}, {Valles}, {Van Hamme}, {Veltz}, {Virtanen}, {Wallut},
  {Wichmann}, {Wilkinson}, {Ziaeepour}, \& {Zschocke}}]{gaia2016}
{Gaia Collaboration}, {Prusti}, T., {de Bruijne}, J.~H.~J., {et~al.} 2016,
  \aap, 595, A1

\bibitem[{{Gaia Collaboration} {et~al.}(2023){Gaia Collaboration}, {Vallenari},
  {Brown}, {Prusti}, {de Bruijne}, {Arenou}, {Babusiaux}, {Biermann},
  {Creevey}, {Ducourant}, {Evans}, {Eyer}, {Guerra}, {Hutton}, {Jordi},
  {Klioner}, {Lammers}, {Lindegren}, {Luri}, {Mignard}, {Panem}, {Pourbaix},
  {Randich}, {Sartoretti}, {Soubiran}, {Tanga}, {Walton}, {Bailer-Jones},
  {Bastian}, {Drimmel}, {Jansen}, {Katz}, {Lattanzi}, {van Leeuwen}, {Bakker},
  {Cacciari}, {Casta{\~n}eda}, {De Angeli}, {Fabricius}, {Fouesneau},
  {Fr{\'e}mat}, {Galluccio}, {Guerrier}, {Heiter}, {Masana}, {Messineo},
  {Mowlavi}, {Nicolas}, {Nienartowicz}, {Pailler}, {Panuzzo}, {Riclet}, {Roux},
  {Seabroke}, {Sordo}, {Th{\'e}venin}, {Gracia-Abril}, {Portell}, {Teyssier},
  {Altmann}, {Andrae}, {Audard}, {Bellas-Velidis}, {Benson}, {Berthier},
  {Blomme}, {Burgess}, {Busonero}, {Busso}, {C{\'a}novas}, {Carry}, {Cellino},
  {Cheek}, {Clementini}, {Damerdji}, {Davidson}, {de Teodoro}, {Nu{\~n}ez
  Campos}, {Delchambre}, {Dell'Oro}, {Esquej}, {Fern{\'a}ndez-Hern{\'a}ndez},
  {Fraile}, {Garabato}, {Garc{\'\i}a-Lario}, {Gosset}, {Haigron}, {Halbwachs},
  {Hambly}, {Harrison}, {Hern{\'a}ndez}, {Hestroffer}, {Hodgkin}, {Holl},
  {Jan{\ss}en}, {Jevardat de Fombelle}, {Jordan}, {Krone-Martins}, {Lanzafame},
  {L{\"o}ffler}, {Marchal}, {Marrese}, {Moitinho}, {Muinonen}, {Osborne},
  {Pancino}, {Pauwels}, {Recio-Blanco}, {Reyl{\'e}}, {Riello}, {Rimoldini},
  {Roegiers}, {Rybizki}, {Sarro}, {Siopis}, {Smith}, {Sozzetti}, {Utrilla},
  {van Leeuwen}, {Abbas}, {{\'A}brah{\'a}m}, {Abreu Aramburu}, {Aerts},
  {Aguado}, {Ajaj}, {Aldea-Montero}, {Altavilla}, {{\'A}lvarez}, {Alves},
  {Anders}, {Anderson}, {Anglada Varela}, {Antoja}, {Baines}, {Baker},
  {Balaguer-N{\'u}{\~n}ez}, {Balbinot}, {Balog}, {Barache}, {Barbato},
  {Barros}, {Barstow}, {Bartolom{\'e}}, {Bassilana}, {Bauchet}, {Becciani},
  {Bellazzini}, {Berihuete}, {Bernet}, {Bertone}, {Bianchi}, {Binnenfeld},
  {Blanco-Cuaresma}, {Blazere}, {Boch}, {Bombrun}, {Bossini}, {Bouquillon},
  {Bragaglia}, {Bramante}, {Breedt}, {Bressan}, {Brouillet}, {Brugaletta},
  {Bucciarelli}, {Burlacu}, {Butkevich}, {Buzzi}, {Caffau}, {Cancelliere},
  {Cantat-Gaudin}, {Carballo}, {Carlucci}, {Carnerero}, {Carrasco},
  {Casamiquela}, {Castellani}, {Castro-Ginard}, {Chaoul}, {Charlot}, {Chemin},
  {Chiaramida}, {Chiavassa}, {Chornay}, {Comoretto}, {Contursi}, {Cooper},
  {Cornez}, {Cowell}, {Crifo}, {Cropper}, {Crosta}, {Crowley}, {Dafonte},
  {Dapergolas}, {David}, {David}, {de Laverny}, {De Luise}, {De March}, {De
  Ridder}, {de Souza}, {de Torres}, {del Peloso}, {del Pozo}, {Delbo},
  {Delgado}, {Delisle}, {Demouchy}, {Dharmawardena}, {Di Matteo}, {Diakite},
  {Diener}, {Distefano}, {Dolding}, {Edvardsson}, {Enke}, {Fabre}, {Fabrizio},
  {Faigler}, {Fedorets}, {Fernique}, {Fienga}, {Figueras}, {Fournier},
  {Fouron}, {Fragkoudi}, {Gai}, {Garcia-Gutierrez}, {Garcia-Reinaldos},
  {Garc{\'\i}a-Torres}, {Garofalo}, {Gavel}, {Gavras}, {Gerlach}, {Geyer},
  {Giacobbe}, {Gilmore}, {Girona}, {Giuffrida}, {Gomel}, {Gomez},
  {Gonz{\'a}lez-N{\'u}{\~n}ez}, {Gonz{\'a}lez-Santamar{\'\i}a},
  {Gonz{\'a}lez-Vidal}, {Granvik}, {Guillout}, {Guiraud},
  {Guti{\'e}rrez-S{\'a}nchez}, {Guy}, {Hatzidimitriou}, {Hauser}, {Haywood},
  {Helmer}, {Helmi}, {Sarmiento}, {Hidalgo}, {Hilger}, {H{\l}adczuk}, {Hobbs},
  {Holland}, {Huckle}, {Jardine}, {Jasniewicz}, {Jean-Antoine Piccolo},
  {Jim{\'e}nez-Arranz}, {Jorissen}, {Juaristi Campillo}, {Julbe}, {Karbevska},
  {Kervella}, {Khanna}, {Kontizas}, {Kordopatis}, {Korn}, {K{\'o}sp{\'a}l},
  {Kostrzewa-Rutkowska}, {Kruszy{\'n}ska}, {Kun}, {Laizeau}, {Lambert},
  {Lanza}, {Lasne}, {Le Campion}, {Lebreton}, {Lebzelter}, {Leccia}, {Leclerc},
  {Lecoeur-Taibi}, {Liao}, {Licata}, {Lindstr{\o}m}, {Lister}, {Livanou},
  {Lobel}, {Lorca}, {Loup}, {Madrero Pardo}, {Magdaleno Romeo}, {Managau},
  {Mann}, {Manteiga}, {Marchant}, {Marconi}, {Marcos}, {Marcos Santos},
  {Mar{\'\i}n Pina}, {Marinoni}, {Marocco}, {Marshall}, {Martin Polo},
  {Mart{\'\i}n-Fleitas}, {Marton}, {Mary}, {Masip}, {Massari},
  {Mastrobuono-Battisti}, {Mazeh}, {McMillan}, {Messina}, {Michalik}, {Millar},
  {Mints}, {Molina}, {Molinaro}, {Moln{\'a}r}, {Monari}, {Mongui{\'o}},
  {Montegriffo}, {Montero}, {Mor}, {Mora}, {Morbidelli}, {Morel}, {Morris},
  {Muraveva}, {Murphy}, {Musella}, {Nagy}, {Noval}, {Oca{\~n}a}, {Ogden},
  {Ordenovic}, {Osinde}, {Pagani}, {Pagano}, {Palaversa}, {Palicio},
  {Pallas-Quintela}, {Panahi}, {Payne-Wardenaar}, {Pe{\~n}alosa Esteller},
  {Penttil{\"a}}, {Pichon}, {Piersimoni}, {Pineau}, {Plachy}, {Plum}, {Poggio},
  {Pr{\v{s}}a}, {Pulone}, {Racero}, {Ragaini}, {Rainer}, {Raiteri}, {Rambaux},
  {Ramos}, {Ramos-Lerate}, {Re Fiorentin}, {Regibo}, {Richards}, {Rios Diaz},
  {Ripepi}, {Riva}, {Rix}, {Rixon}, {Robichon}, {Robin}, {Robin}, {Roelens},
  {Rogues}, {Rohrbasser}, {Romero-G{\'o}mez}, {Rowell}, {Royer}, {Ruz Mieres},
  {Rybicki}, {Sadowski}, {S{\'a}ez N{\'u}{\~n}ez}, {Sagrist{\`a} Sell{\'e}s},
  {Sahlmann}, {Salguero}, {Samaras}, {Sanchez Gimenez}, {Sanna},
  {Santove{\~n}a}, {Sarasso}, {Schultheis}, {Sciacca}, {Segol}, {Segovia},
  {S{\'e}gransan}, {Semeux}, {Shahaf}, {Siddiqui}, {Siebert}, {Siltala},
  {Silvelo}, {Slezak}, {Slezak}, {Smart}, {Snaith}, {Solano}, {Solitro},
  {Souami}, {Souchay}, {Spagna}, {Spina}, {Spoto}, {Steele},
  {Steidelm{\"u}ller}, {Stephenson}, {S{\"u}veges}, {Surdej}, {Szabados},
  {Szegedi-Elek}, {Taris}, {Taylor}, {Teixeira}, {Tolomei}, {Tonello}, {Torra},
  {Torra}, {Torralba Elipe}, {Trabucchi}, {Tsounis}, {Turon}, {Ulla}, {Unger},
  {Vaillant}, {van Dillen}, {van Reeven}, {Vanel}, {Vecchiato}, {Viala},
  {Vicente}, {Voutsinas}, {Weiler}, {Wevers}, {Wyrzykowski}, {Yoldas}, {Yvard},
  {Zhao}, {Zorec}, {Zucker}, \& {Zwitter}}]{brown2023gaiadr3}
{Gaia Collaboration}, {Vallenari}, A., {Brown}, A.~G.~A., {et~al.} 2023, \aap,
  674, A1

\bibitem[{G{\"u}nther \& Daylan(2021)}]{gunther2021allesfitter}
G{\"u}nther, M.~N., \& Daylan, T. 2021, \apjs, 254, 13

\bibitem[{Hamer \& Schlaufman(2022)}]{hamer2022evidence}
Hamer, J.~H., \& Schlaufman, K.~C. 2022, arXiv preprint arXiv:2205.00040

\bibitem[{Harre {et~al.}(2023)Harre, Smith, Hirano, Csizmadia, Triaud, \&
  Anderson}]{harre2023wasp106b}
Harre, J.-V., Smith, A. M.~S., Hirano, T., {et~al.} 2023, The Astronomical
  Journal, 166, 159.
\newblock \url{https://dx.doi.org/10.3847/1538-3881/acf46d}

\bibitem[{Harris {et~al.}(2020)Harris, Millman, van~der Walt, Gommers,
  Virtanen, Cournapeau, Wieser, Taylor, Berg, Smith,
  {et~al.}}]{harris2020array}
Harris, C.~R., Millman, K.~J., van~der Walt, S.~J., {et~al.} 2020, Nature, 585,
  357

\bibitem[{{H{\'e}brard} {et~al.}(2008){H{\'e}brard}, {Bouchy}, {Pont},
  {Loeillet}, {Rabus}, {Bonfils}, {Moutou}, {Boisse}, {Delfosse}, {Desort},
  {Eggenberger}, {Ehrenreich}, {Forveille}, {Lagrange}, {Lovis}, {Mayor},
  {Pepe}, {Perrier}, {Queloz}, {Santos}, {S{\'e}gransan}, {Udry}, \&
  {Vidal-Madjar}}]{hebrard2008misaligned}
{H{\'e}brard}, G., {Bouchy}, F., {Pont}, F., {et~al.} 2008, \aap, 488, 763

\bibitem[{{Hirano} {et~al.}(2011){Hirano}, {Narita}, {Sato}, {Winn}, {Aoki},
  {Tamura}, {Taruya}, \& {Suto}}]{hirano2011further}
{Hirano}, T., {Narita}, N., {Sato}, B., {et~al.} 2011, \pasj, 63, L57

\bibitem[{{Hixenbaugh} {et~al.}(2023){Hixenbaugh}, {Wang}, {Rice}, \&
  {Wang}}]{hixenbaugh2023spin}
{Hixenbaugh}, K., {Wang}, X.-Y., {Rice}, M., \& {Wang}, S. 2023, \apjl, 949,
  L35

\bibitem[{Huang {et~al.}(2016)Huang, Wu, \& Triaud}]{huang2016warm}
Huang, C., Wu, Y., \& Triaud, A. H. M.~J. 2016, \apj, 825, 98

\bibitem[{Hunter(2007)}]{hunter2007matplotlib}
Hunter, J.~D. 2007, Computing in science \& engineering, 9, 90

\bibitem[{{Jensen}(2013)}]{Jensen:2013}
{Jensen}, E. 2013, {Tapir: A web interface for transit/eclipse observability},
  , , ascl:1306.007

\bibitem[{Jordán {et~al.}(2020)Jordán, Brahm, Espinoza, Henning, Jones,
  Kossakowski, Sarkis, Trifonov, Rojas, Torres, Drass, Nandakumar, Barbieri,
  Davis, Wang, Bayliss, Bouma, Dragomir, Eastman, Daylan, Guerrero, Barclay,
  Ting, Henze, Ricker, Vanderspek, Latham, Seager, Winn, Jenkins, Wittenmyer,
  Bowler, Crossfield, Horner, Kane, Kielkopf, Morton, Plavchan, Tinney,
  Addison, Mengel, Okumura, Shahaf, Mazeh, Rabus, Shporer, Ziegler, Mann, \&
  Hart}]{Jordan_2020}
Jordán, A., Brahm, R., Espinoza, N., {et~al.} 2020, The Astronomical Journal,
  159, 145.
\newblock \url{https://dx.doi.org/10.3847/1538-3881/ab6f67}

\bibitem[{{Kipping}(2013)}]{kipping2013ld}
{Kipping}, D.~M. 2013, \mnras, 435, 2152

\bibitem[{{Kley} \& {Dirksen}(2006)}]{kley2006disk}
{Kley}, W., \& {Dirksen}, G. 2006, \aap, 447, 369

\bibitem[{Kozai(1962)}]{kozai1962secular}
Kozai, Y. 1962, \aj, 67, 591

\bibitem[{Kraft(1967)}]{kraft1967studies}
Kraft, R.~P. 1967, \apj, 150, 551

\bibitem[{Li \& Lai(2023)}]{Li2023eccentricdisk}
Li, J., \& Lai, D. 2023, The Astrophysical Journal, 956, 17.
\newblock \url{https://dx.doi.org/10.3847/1538-4357/aced89}

\bibitem[{Lidov(1962)}]{lidov1962evolution}
Lidov, M.~L. 1962, Planetary and Space Science, 9, 719

\bibitem[{{Lubin} {et~al.}(2023){Lubin}, {Wang}, {Rice}, {Dong}, {Wang},
  {Radzom}, {Robertson}, {Stefansson}, {Alvarado-Montes}, {Beard}, {Bender},
  {Gupta}, {Halverson}, {Kanodia}, {Li}, {Lin}, {Logsdon}, {Lubar},
  {Mahadevan}, {Ninan}, {Rajagopal}, {Roy}, {Schwab}, \&
  {Wright}}]{lubin2023toi}
{Lubin}, J., {Wang}, X.-Y., {Rice}, M., {et~al.} 2023, \apjl, 959, L5

\bibitem[{{MAST Team}(2021)}]{tess_all_sectors}
{MAST Team}. 2021, TESS Light Curves - All Sectors,  STScI/MAST,
  doi:10.17909/T9-NMC8-F686.
\newblock
  \url{http://archive.stsci.edu/doi/resolve/resolve.html?doi=10.17909/t9-nmc8-f686}

\bibitem[{{Matsakos} \& {K{\"o}nigl}(2017)}]{matsakos2017disk}
{Matsakos}, T., \& {K{\"o}nigl}, A. 2017, \aj, 153, 60

\bibitem[{Maxted(2016)}]{maxted2016ellc}
Maxted, P. 2016, Astronomy \& Astrophysics, 591, A111

\bibitem[{{McCully} {et~al.}(2018){McCully}, {Volgenau}, {Harbeck}, {Lister},
  {Saunders}, {Turner}, {Siiverd}, \& {Bowman}}]{McCully:2018}
{McCully}, C., {Volgenau}, N.~H., {Harbeck}, D.-R., {et~al.} 2018, in Society
  of Photo-Optical Instrumentation Engineers (SPIE) Conference Series, Vol.
  10707, \procspie, 107070K

\bibitem[{McKinney(2010)}]{mckinney2010data}
McKinney, W. 2010, in Proceedings of the 9th Python in Science Conference, Vol.
  445, Austin, TX, 51--56

\bibitem[{{McLaughlin}(1924)}]{McLaughlin1924}
{McLaughlin}, D.~B. 1924, \apj, 60, 22

\bibitem[{{Morgan} {et~al.}(2023){Morgan}, {Bowler}, {Tran}, {Petigura},
  {Nagpal}, \& {Blunt}}]{morgan2023signs}
{Morgan}, M., {Bowler}, B.~P., {Tran}, Q.~H., {et~al.} 2023, arXiv e-prints,
  arXiv:2310.18445

\bibitem[{{Murray} {et~al.}(2022){Murray}, {Hadden}, \&
  {Holman}}]{murray2022effects}
{Murray}, Z., {Hadden}, S., \& {Holman}, M.~J. 2022, \apj, 931, 66

\bibitem[{Naoz(2016)}]{naoz2016eccentric}
Naoz, S. 2016, \araa, 54, 441

\bibitem[{Naoz {et~al.}(2011)Naoz, Farr, Lithwick, Rasio, \&
  Teyssandier}]{naoz2011hot}
Naoz, S., Farr, W.~M., Lithwick, Y., Rasio, F.~A., \& Teyssandier, J. 2011,
  Nature, 473, 187

\bibitem[{Naoz {et~al.}(2012)Naoz, Farr, \& Rasio}]{naoz2012formation}
Naoz, S., Farr, W.~M., \& Rasio, F.~A. 2012, \apjl, 754, L36

\bibitem[{{NASA Exoplanet Archive}(2023)}]{ps_nasaexo}
{NASA Exoplanet Archive}. 2023, Planetary Systems, vVersion: 2023-10-20 HH:MM,
  NExScI-Caltech/IPAC, doi:10.26133/NEA12.
\newblock \url{https://catcopy.ipac.caltech.edu/dois/doi.php?id=10.26133/NEA12}

\bibitem[{{Neveu-VanMalle} {et~al.}(2014){Neveu-VanMalle}, {Queloz},
  {Anderson}, {Charbonnel}, {Collier Cameron}, {Delrez}, {Gillon}, {Hellier},
  {Jehin}, {Lendl}, {Maxted}, {Pepe}, {Pollacco}, {S{\'e}gransan}, {Smalley},
  {Smith}, {Southworth}, {Triaud}, {Udry}, \& {West}}]{neveu2014wasp}
{Neveu-VanMalle}, M., {Queloz}, D., {Anderson}, D.~R., {et~al.} 2014, \aap,
  572, A49

\bibitem[{Ogilvie(2014)}]{ogilvie2014tidal}
Ogilvie, G.~I. 2014, \araa, 52, 171

\bibitem[{Oliphant(2006)}]{oliphant2006guide}
Oliphant, T.~E. 2006, A guide to NumPy, Vol.~1 (Trelgol Publishing USA)

\bibitem[{Pepe {et~al.}(2021)Pepe, Cristiani, Rebolo, Santos, Dekker, Cabral,
  Marcantonio, Figueira, Curto, Lovis, Mayor, M{\'e}gevand, Molaro, Riva,
  Osorio, Amate, Manescau, Pasquini, Zerbi, Adibekyan, Abreu, Affolter,
  Alibert, Aliverti, Allart, Prieto, {\'A}lvarez, Alves, Avila, Baldini, Bandy,
  Barros, Benz, Bianco, Borsa, Bourrier, Bouchy, Broeg, Calderone, Cirami,
  Coelho, Conconi, Coretti, Cumani, Cupani, D'Odorico, Damasso, Deiries,
  Delabre, Demangeon, Dumusque, Ehrenreich, Faria, Fragoso, Genolet, Genoni,
  Santos, Hern{\'a}ndez, Hughes, Iwert, Kerber, Knudstrup, Landoni, Lavie,
  {Lillo-Box}, Lizon, Maire, Martins, Mehner, Micela, Modigliani, Monteiro,
  Monteiro, Moschetti, Murphy, Nunes, Oggioni, Oliveira, Oshagh, Pall{\'e},
  Pariani, Poretti, Rasilla, Rebord{\~a}o, Redaelli, Tschudi, Santin, Santos,
  S{\'e}gransan, Schmidt, Segovia, Sosnowska, Sozzetti, Sousa, Span{\`o},
  Mascare{\~n}o, Tabernero, Tenegi, Udry, \& Zanutta}]{pepe2021espresso}
Pepe, F., Cristiani, S., Rebolo, R., {et~al.} 2021, Astronomy \& Astrophysics,
  645, A96

\bibitem[{Perryman {et~al.}(2014)Perryman, Hartman, Bakos, \&
  Lindegren}]{perryman2014gaia}
Perryman, M., Hartman, J., Bakos, G.~A., \& Lindegren, L. 2014, The
  Astrophysical Journal, 797, 14.
\newblock \url{https://dx.doi.org/10.1088/0004-637X/797/1/14}

\bibitem[{Petrovich(2015)}]{petrovich2015CHEM}
Petrovich, C. 2015, The Astrophysical Journal, 805, 75.
\newblock \url{https://dx.doi.org/10.1088/0004-637X/805/1/75}

\bibitem[{Petrovich \& Tremaine(2016)}]{petrovich2016warm}
Petrovich, C., \& Tremaine, S. 2016, \apj, 829, 132

\bibitem[{Petrovich {et~al.}(2014)Petrovich, Tremaine, \&
  Rafikov}]{petrovich2014cicularwj}
Petrovich, C., Tremaine, S., \& Rafikov, R. 2014, The Astrophysical Journal,
  786, 101.
\newblock \url{https://dx.doi.org/10.1088/0004-637X/786/2/101}

\bibitem[{{Queloz} {et~al.}(2000){Queloz}, {Eggenberger}, {Mayor}, {Perrier},
  {Beuzit}, {Naef}, {Sivan}, \& {Udry}}]{RM2000}
{Queloz}, D., {Eggenberger}, A., {Mayor}, M., {et~al.} 2000, \aap, 359, L13

\bibitem[{{Rice} {et~al.}(2023{\natexlab{a}}){Rice}, {Wang}, {Gerbig}, {Wang},
  {Dai}, {Tyler}, {Isaacson}, \& {Howard}}]{rice2023qatar6}
{Rice}, M., {Wang}, S., {Gerbig}, K., {et~al.} 2023{\natexlab{a}}, \aj, 165, 65

\bibitem[{{Rice} {et~al.}(2022){Rice}, {Wang}, \&
  {Laughlin}}]{HJ_obliquity_Rice_2022}
{Rice}, M., {Wang}, S., \& {Laughlin}, G. 2022, \apjl, 926, L17

\bibitem[{Rice {et~al.}(2021)Rice, Wang, Howard, Isaacson, Dai, Wang, Beard,
  Behmard, Brinkman, Rubenzahl, {et~al.}}]{rice2021soles}
Rice, M., Wang, S., Howard, A.~W., {et~al.} 2021, \aj, 162, 182

\bibitem[{Rice {et~al.}(2022{\natexlab{a}})Rice, Wang, Wang, Stefánsson,
  Isaacson, Howard, Logsdon, Schweiker, Dai, Brinkman, Giacalone, \&
  Holcomb}]{WJ_obliquity_Rice_2022}
Rice, M., Wang, S., Wang, X.-Y., {et~al.} 2022{\natexlab{a}}, The Astronomical
  Journal, 164, 104.
\newblock \url{https://dx.doi.org/10.3847/1538-3881/ac8153}

\bibitem[{Rice {et~al.}(2022{\natexlab{b}})Rice, Wang, Wang, Stef{\'a}nsson,
  Isaacson, Howard, Logsdon, Schweiker, Dai, Brinkman,
  {et~al.}}]{rice2022tendency}
---. 2022{\natexlab{b}}, \aj, 164, 104

\bibitem[{{Rice} {et~al.}(2023{\natexlab{b}}){Rice}, {Wang}, {Wang}, {Shporer},
  {Barkaoui}, {Brahm}, {Collins}, {Jord{\'a}n}, {Lowson}, {Butler}, {Crane},
  {Shectman}, {Teske}, {Osip}, {Collins}, {Murgas}, {Boyle}, {Pozuelos},
  {Timmermans}, {Jehin}, \& {Gillon}}]{rice2023evidence}
{Rice}, M., {Wang}, X.-Y., {Wang}, S., {et~al.} 2023{\natexlab{b}}, \aj, 166,
  266

\bibitem[{{Ricker} {et~al.}(2015){Ricker}, {Winn}, {Vanderspek}, {Latham},
  {Bakos}, {Bean}, {Berta-Thompson}, {Brown}, {Buchhave}, {Butler}, {Butler},
  {Chaplin}, {Charbonneau}, {Christensen-Dalsgaard}, {Clampin}, {Deming},
  {Doty}, {De Lee}, {Dressing}, {Dunham}, {Endl}, {Fressin}, {Ge}, {Henning},
  {Holman}, {Howard}, {Ida}, {Jenkins}, {Jernigan}, {Johnson}, {Kaltenegger},
  {Kawai}, {Kjeldsen}, {Laughlin}, {Levine}, {Lin}, {Lissauer}, {MacQueen},
  {Marcy}, {McCullough}, {Morton}, {Narita}, {Paegert}, {Palle}, {Pepe},
  {Pepper}, {Quirrenbach}, {Rinehart}, {Sasselov}, {Sato}, {Seager},
  {Sozzetti}, {Stassun}, {Sullivan}, {Szentgyorgyi}, {Torres}, {Udry}, \&
  {Villasenor}}]{ricker2015tess}
{Ricker}, G.~R., {Winn}, J.~N., {Vanderspek}, R., {et~al.} 2015, Journal of
  Astronomical Telescopes, Instruments, and Systems, 1, 014003

\bibitem[{Rogers {et~al.}(2012)Rogers, Lin, \& Lau}]{rogers2012internal}
Rogers, T., Lin, D.~N., \& Lau, H. H.~B. 2012, \apjl, 758, L6

\bibitem[{{Rogers} {et~al.}(2013){Rogers}, {Lin}, {McElwaine}, \&
  {Lau}}]{rogers2013igw}
{Rogers}, T.~M., {Lin}, D.~N.~C., {McElwaine}, J.~N., \& {Lau}, H.~H.~B. 2013,
  \apj, 772, 21

\bibitem[{{Romanova} {et~al.}(2021){Romanova}, {Koldoba}, {Ustyugova},
  {Blinova}, {Lai}, \& {Lovelace}}]{romanova2021MHD}
{Romanova}, M.~M., {Koldoba}, A.~V., {Ustyugova}, G.~V., {et~al.} 2021, \mnras,
  506, 372

\bibitem[{Romanova {et~al.}(2023)Romanova, Koldoba, Ustyugova, Lai, \&
  Lovelace}]{romanova2023higheindisk}
Romanova, M.~M., Koldoba, A.~V., Ustyugova, G.~V., Lai, D., \& Lovelace, R.
  V.~E. 2023, Monthly Notices of the Royal Astronomical Society, 523, 2832.
\newblock \url{https://doi.org/10.1093/mnras/stad987}

\bibitem[{{Rossiter}(1924)}]{Rossiter1924}
{Rossiter}, R.~A. 1924, \apj, 60, 15

\bibitem[{{Schlaufman}(2010)}]{Schlaufman_2010}
{Schlaufman}, K.~C. 2010, \apj, 719, 602

\bibitem[{Sedaghati {et~al.}(2023)Sedaghati, Jordán, Brahm, Muñoz, Petrovich,
  \& Hobson}]{TOI677_Sedaghati_2023}
Sedaghati, E., Jordán, A., Brahm, R., {et~al.} 2023, The Astronomical Journal,
  166, 130.
\newblock \url{https://dx.doi.org/10.3847/1538-3881/acea84}

\bibitem[{Souami \& Souchay(2012)}]{souami2012solar}
Souami, D., \& Souchay, J. 2012, Astronomy \& Astrophysics, 543, A133

\bibitem[{Southworth(2011)}]{southworth2011homogeneous}
Southworth, J. 2011, \mnras, 417, 2166

\bibitem[{{Takaishi} {et~al.}(2020){Takaishi}, {Tsukamoto}, \&
  {Suto}}]{takaishi2020diskalign}
{Takaishi}, D., {Tsukamoto}, Y., \& {Suto}, Y. 2020, \mnras, 492, 5641

\bibitem[{{Temple} {et~al.}(2017){Temple}, {Hellier}, {Albrow}, {Anderson},
  {Bayliss}, {Beatty}, {Bieryla}, {Brown}, {Cargile}, {Collier Cameron},
  {Collins}, {Col{\'o}n}, {Curtis}, {D'Ago}, {Delrez}, {Eastman}, {Gaudi},
  {Gillon}, {Gregorio}, {James}, {Jehin}, {Joner}, {Kielkopf}, {Kuhn},
  {Labadie-Bartz}, {Latham}, {Lendl}, {Lund}, {Malpas}, {Maxted}, {Myers},
  {Oberst}, {Pepe}, {Pepper}, {Pollacco}, {Queloz}, {Rodriguez},
  {S{\'e}gransan}, {Siverd}, {Smalley}, {Stassun}, {Stevens}, {Stockdale},
  {Tan}, {Triaud}, {Udry}, {Villanueva}, {West}, \& {Zhou}}]{temple2017wasp}
{Temple}, L.~Y., {Hellier}, C., {Albrow}, M.~D., {et~al.} 2017, \mnras, 471,
  2743

\bibitem[{{Temple} {et~al.}(2019){Temple}, {Hellier}, {Anderson}, {Barkaoui},
  {Bouchy}, {Brown}, {Burdanov}, {Collier Cameron}, {Delrez}, {Ducrot},
  {Evans}, {Gillon}, {Jehin}, {Lendl}, {Maxted}, {McCormac}, {Murray},
  {Nielsen}, {Pepe}, {Pollacco}, {Queloz}, {S{\'e}gransan}, {Smalley},
  {Thompson}, {Triaud}, {Turner}, {Udry}, {West}, \&
  {Zouhair}}]{temple2019wasp}
{Temple}, L.~Y., {Hellier}, C., {Anderson}, D.~R., {et~al.} 2019, \mnras, 490,
  2467

\bibitem[{Teyssandier {et~al.}(2019)Teyssandier, Lai, \&
  Vick}]{teyssandier2019formation}
Teyssandier, J., Lai, D., \& Vick, M. 2019, \mnras, 486, 2265

\bibitem[{Tremaine(2023)}]{tremaine2023dynamics}
Tremaine, S. 2023, Dynamics of Planetary Systems, Vol.~63 (Princeton University
  Press)

\bibitem[{{Triaud, A. H. M. J.} {et~al.}(2010){Triaud, A. H. M. J.}, {Collier
  Cameron, A.}, {Queloz, D.}, {Anderson, D. R.}, {Gillon, M.}, {Hebb, L.},
  {Hellier, C.}, {Loeillet, B.}, {Maxted, P. F. L.}, {Mayor, M.}, {Pepe, F.},
  {Pollacco, D.}, {Ségransan, D.}, {Smalley, B.}, {Udry, S.}, {West, R. G.},
  \& {Wheatley, P. J.}}]{triaud2010hj}
{Triaud, A. H. M. J.}, {Collier Cameron, A.}, {Queloz, D.}, {et~al.} 2010,
  A\&A, 524, A25.
\newblock \url{https://doi.org/10.1051/0004-6361/201014525}

\bibitem[{Virtanen {et~al.}(2020)Virtanen, Gommers, Oliphant, Haberland, Reddy,
  Cournapeau, Burovski, Peterson, Weckesser, Bright,
  {et~al.}}]{virtanen2020scipy}
Virtanen, P., Gommers, R., Oliphant, T.~E., {et~al.} 2020, Nature Methods, 17,
  261

\bibitem[{{von Zeipel}(1910)}]{von_zeipel_1910}
{von Zeipel}, H. 1910, Astronomische Nachrichten, 183, 345

\bibitem[{Walt {et~al.}(2011)Walt, Colbert, \& Varoquaux}]{walt2011numpy}
Walt, S. v.~d., Colbert, S.~C., \& Varoquaux, G. 2011, Computing in Science \&
  Engineering, 13, 22

\bibitem[{Wang {et~al.}(2022)Wang, Rice, Wang, Pu, Stef{\'a}nsson, Mahadevan,
  Radzom, Giacalone, Wu, Esposito, {et~al.}}]{wang2022aligned}
Wang, X.-Y., Rice, M., Wang, S., {et~al.} 2022, \apjl, 926, L8

\bibitem[{{Watanabe} {et~al.}(2022){Watanabe}, {Narita}, {Palle}, {Fukui},
  {Kusakabe}, {Parviainen}, {Murgas}, {Casasayas-Barris}, {Johnson}, {Sato},
  {Livingston}, {de Leon}, {Mori}, {Nishiumi}, {Terada}, {Esparza-Borges}, \&
  {Kawauchi}}]{watanabe2022nodal}
{Watanabe}, N., {Narita}, N., {Palle}, E., {et~al.} 2022, \mnras, 512, 4404

\bibitem[{Winn {et~al.}(2010)Winn, Fabrycky, Albrecht, \&
  Johnson}]{winn2010hot}
Winn, J.~N., Fabrycky, D., Albrecht, S., \& Johnson, J.~A. 2010, \apjl, 718,
  L145

\bibitem[{{Winn} {et~al.}(2010){Winn}, {Fabrycky}, {Albrecht}, \&
  {Johnson}}]{Winn_2010}
{Winn}, J.~N., {Fabrycky}, D., {Albrecht}, S., \& {Johnson}, J.~A. 2010, \apjl,
  718, L145

\bibitem[{Winn \& Fabrycky(2015)}]{HJ_ARAA2015}
Winn, J.~N., \& Fabrycky, D.~C. 2015, Annual Review of Astronomy and
  Astrophysics, 53, 409

\bibitem[{{Winn} {et~al.}(2009){Winn}, {Johnson}, {Fabrycky}, {Howard},
  {Marcy}, {Narita}, {Crossfield}, {Suto}, {Turner}, {Esquerdo}, \&
  {Holman}}]{winn2009on}
{Winn}, J.~N., {Johnson}, J.~A., {Fabrycky}, D., {et~al.} 2009, \apj, 700, 302

\bibitem[{Worku {et~al.}(2022)Worku, Wang, Burt, Rice, Wang, Wang, Vogt,
  Butler, Addison, Holden, {et~al.}}]{worku2022revisiting}
Worku, K., Wang, S., Burt, J., {et~al.} 2022, \aj, 163, 158

\bibitem[{{Wright} {et~al.}(2023){Wright}, {Rice}, {Wang}, {Hixenbaugh}, \&
  {Wang}}]{wright2023soles}
{Wright}, J., {Rice}, M., {Wang}, X.-Y., {Hixenbaugh}, K., \& {Wang}, S. 2023,
  \aj, 166, 217

\bibitem[{Wu {et~al.}(2023)Wu, Rice, \& Wang}]{Wu_2023}
Wu, D.-H., Rice, M., \& Wang, S. 2023, The Astronomical Journal, 165, 171.
\newblock \url{https://dx.doi.org/10.3847/1538-3881/acbf3f}

\bibitem[{Wu \& Murray(2003)}]{wu2003planet}
Wu, Y., \& Murray, N. 2003, \apj, 589, 605

\bibitem[{Yee {et~al.}(2018)Yee, Petigura, Fulton, Knutson, Batygin, Bakos,
  Hartman, Hirsch, Howard, Isaacson, Kosiarek, Sinukoff, \&
  Weiss}]{yee2018hatp11}
Yee, S.~W., Petigura, E.~A., Fulton, B.~J., {et~al.} 2018, The Astronomical
  Journal, 155, 255.
\newblock \url{https://dx.doi.org/10.3847/1538-3881/aabfec}

\bibitem[{Yu {et~al.}(2018)Yu, Zhou, Rodriguez, Huang, Vanderburg, Quinn,
  Gaudi, Beichman, Berlind, Bieryla, {et~al.}}]{yu2018epic}
Yu, L., Zhou, G., Rodriguez, J.~E., {et~al.} 2018, \aj, 156, 250

\bibitem[{Zhou {et~al.}(2020)Zhou, Winn, Newton, Quinn, Rodriguez, Mann,
  Rizzuto, Vanderburg, Huang, Latham, Teske, Wang, Shectman, Butler, Crane,
  Thompson, Henry, Paredes, Jao, James, \& Hinojosa}]{zhou2020obliquity}
Zhou, G., Winn, J.~N., Newton, E.~R., {et~al.} 2020, The Astrophysical Journal
  Letters, 892, L21.
\newblock \url{https://dx.doi.org/10.3847/2041-8213/ab7d3c}

\bibitem[{Zink \& Howard(2023)}]{zink2023hot}
Zink, J.~K., \& Howard, A.~W. 2023, The Astrophysical Journal Letters, 956, L29

\end{thebibliography}
\bibliographystyle{aasjournal}

\end{document}